\documentclass[reprint,amsmath,amssymb,aps,prb,amssymb,superscriptaddress]{revtex4-1}
\usepackage[utf8]{inputenc}
\usepackage{color}
\usepackage{amsmath}
\usepackage{mathtools}
\usepackage{bm}
\usepackage{amsfonts}
\usepackage{dcolumn}
\usepackage{natbib}
\usepackage{bbold}
\usepackage{soul}
\usepackage[dvipsnames]{xcolor}
\usepackage{amssymb}
\usepackage[breaklinks=true,colorlinks,citecolor=blue,linkcolor=blue,urlcolor=blue]{hyperref}
\usepackage{graphicx}
\def\be{\begin{equation}}
\def\ee{\end{equation}}
\def \bea{\begin{eqnarray}}
\def \eea{\end{eqnarray}}
\def \nn{\nonumber}

\begin{document}
\title{Intrinsic nonreciprocal bulk plasmons in noncentrosymmetric magnetic systems}
	\author{Debasis Dutta$^\parallel$}
	\email{ddebasis@iitk.ac.in}
	\author{Atasi Chakraborty$^\parallel$}
	\email{atasic@iitk.ac.in}
	\author{Amit Agarwal}
	\email{amitag@iitk.ac.in}
	\thanks{\\$\dagger$ Corresponding author\\
     \noindent$\parallel$ These authors contributed equally to this work.}
	\affiliation{Department of Physics, Indian Institute of Technology Kanpur, Kanpur-208016, India}
\date{\today}

\begin{abstract}
{Nonreciprocal plasmonics plays a crucial role in enabling one-way light propagation at the nanoscale and is a fundamental building block for photonic applications. Here, we investigate \textit{intrinsic} nonreciprocity in bulk plasmon dispersion in systems that break both parity and time-reversal symmetry. We demonstrate that both interband and intraband bulk plasmon modes exhibit \textit{intrinsically} asymmetric dispersion depending on the sign of wavevector.}
Our study reveals that the \textit{intrinsic} nonreciprocity in interband plasmon dispersion is governed by quantum metric connection.
The nonreciprocity in the intraband plasmon dispersion is dictated by the quantum metric dipole and a higher-order `Drude' weight-like term. We corroborate our findings via explicit numerical calculations for the two-dimensional Qi-Wu-Zhang model and demonstrate the existence of \textit{intrinsic} nonreciprocal intraband and interband plasmon modes in moir\'{e} systems such as twisted bilayer graphene. {Our findings offer new insights into the underlying physics of nonreciprocal plasmonics and pave way for designing novel photonic devices.} 
\end{abstract}

\maketitle

\section{Introduction}

{Plasmons are the collective density oscillation modes of a charged fluid. In electron liquids in metallic systems, plasmons provide an efficient way of coupling electromagnetic waves with electrons via surface-plasmon polariton at the nanoscale~\cite{Koppens2015A,Jablan2009, FrankHKoppen2021B, ZAYATS2005, Agarwal2018, carlo2021, Politano_t_2018}. Plasmon allows manipulating light beyond the diffraction limit~\cite{DNBasov2016}. For an unbiased reciprocal plasmonic medium, the plasmon dispersion does not depend on the sign of the wave vector $\bm{q}$ {\it i.e.}, $\omega_p(\bm{q})=\omega_p(-\bm{q})$. Reciprocity is an intrinsic property of a linear and time-reversal invariant photonic system, which allows symmetric propagation of forward and backward-moving light~\cite{Caloz2018}. In optics, this is known as the Lorentz reciprocity principle and is intimately related to the invariance
of Maxwell’s equations under time-reversal symmetry~\cite{RJPotton_2004}. 
So, the breakdown of time-reversal symmetry is essential to observe a nonreciprocal or asymmetric light propagation. In addition to this, all nonreciprocal phenomena occur in systems with broken inversion symmetry~\cite{Tokura2018N}. 
The breakdown of reciprocity produces an asymmetry between forward and backward propagating density oscillations for the same magnitude of the wave vector giving rise to nonreciprocal plasmons with $\omega_p(\bm{q})\neq  \omega_p(-\bm{q})$. These nonreciprocal plasmons are promising for photonics applications like optical communication~\cite{Chin2013, Guddala2021}, oneway waveguiding~\cite{Yu_2012}, optical transmission~\cite{PRL2010Khanikeav}, and optical isolators~\cite{AIP2009} among others.}

Generally, nonreciprocity in the plasmon dispersion is induced i) through the plasmonic doppler effect via an externally applied drift electric current~\cite{DNBasov2021, Lewandowski2020}, or ii) in the presence of a static magnetic field~\cite{Yu2008, Yu_2012, Norio2021, Fetter1985, Fetter1986, LianFu2016}. 
Additionally, chiral edge plasmon has been proposed in a two-dimensional gapped Dirac material illuminated with circularly polarized light~\cite{TonyLow2016}, and at the edges of anomalous Hall metals~\cite{Song2015}. 
{In the absence of inversion (${\mathcal P}$) and time-reversal (${\mathcal T}$) symmetry, quantum materials are known to exhibit different \textit{intrinsic} nonreciprocal responses dictated by different band-geometric quantities~\cite{Tokura2018N, Justin_song2022}. }  

Recently, \textit{intrinsic} nonreciprocal bulk intraband plasmon has been explored in  
${\mathcal P}$ and ${\mathcal T}$ symmetry broken systems~\cite{Justin_song2022}.
Motivated by this, we develop a theory of \textit{intrinsic} nonreciprocity in interband plasmons in this paper. 
{An interband plasmon is formed by the Coulomb interaction-induced collective oscillation of electrons undergoing interband transitions between the nested bands having singularity in the electron-hole joint density of states (JDOS) spectrum~\cite{Atasi_2022,FrankHLKoppen2021}.} 
Here, we demonstrate that the interband plasmon becomes \textit{intrinsically} nonreciprocal when both ${\mathcal P}$ and ${\mathcal T}$ symmetries of the system are broken. 
We calculate the asymmetry of the interband plasmon dispersion, $\delta \omega_{p}(|\bm{q}|)\equiv\omega_p(\bm{q})-\omega_p(-\bm{q})$, to be  
\bea
\delta \omega_{p}^{\rm inter}(|\bm{q}|) \sim q_aq_bq_cV_{|\bm{q}|}^{(d)}\sum_{\bm{k}}\sum_{s,s^{\prime}}^{s^{\prime}\ne s} f_{s,\bm{k}}\Gamma^{abc}_{s^{\prime}s}(\bm{k})~, 
\label{Eq1_inter}
\eea
 {where $a$, $b$, $c$ labels Cartesian directions, and $s$, $s^{\prime}$ denotes band indices.}
Here, $\Gamma^{abc}_{s^{\prime}s}(\bm{k})$ represents the quantum metric connection~\cite{AhnPRX2020,Agarwal2022A} {defined in  Eq.~\eqref{Eq_L1}}, $V_{|\bm{q}|}^{(d)}$ is the Fourier transform of Coulomb interaction in $d$- dimension,  $f_{s,\bm{k}}$ denotes Fermi-Dirac distribution function, and sum over repeated indices is implied. 
The $\bm k$-integral in Eq.~\eqref{Eq1_inter} is non-zero only when the system lacks both $\cal P$ and $\cal T$ symmetries. This nonreciprocity in the interband plasmon dispersion is dictated by the quantum metric connection, which depends on the quantum geometry of Bloch states. 

In addition to demonstrating nonreciprocity in the interband plasmon modes, we revisit the theory of \textit{intrinsic} nonreciprocal intraband plasmon. We show that for simultaneous  ${\cal P}$ and ${\cal T}$ broken systems, the asymmetry  
of intraband plasmon dispersion can be approximated as 
\bea
\delta \omega_{p}^{\rm intra}(|\bm{q}|) \approx V_{|\bm{q}|}^{(d)}{\mathcal Q}_{abc}q_aq_bq_c + \frac{{\mathcal C}_{abc}}{{\mathcal D}_{mn}}\frac{q_aq_bq_c}{q_mq_n}~,
\label{Eq2_intra}
\eea
in the long wavelength limit.  
Here, ${\mathcal Q}_{abc}$ represents the quantum metric dipole, ${\mathcal D}_{ab}$ is Drude weight and ${\mathcal C}_{abc}$ represents higher-order Drude weight like term~\cite{Xiao_PRL2019,Hughes2019,Atasi_2022}. {These quantities are explicitly defined below Eq.~\eqref{Eq_A1}, Eq.~\eqref{Eq_A2}, and Eq.~\eqref{A3}, respectively.}
We explicitly demonstrate the existence of these nonreciprocal plasmons in the Qu-Wu-Zhang (QWZ) model, and in moir\'{e} superlattices such as twisted bilayer graphene (TBG).

Our manuscript is organized as follows: In section~\ref{Sec-II}, we formulate the generic theory of nonreciprocity for intraband plasmon by calculating the density-density response function and the dynamical dielectric function within random phase approximation. In section~\ref{section-III}, we formulate the general theory of \textit{intrinsic} nonreciprocal interband plasmons. We show the existence of nonreciprocal intraband and interband plasmon modes in ${\mathcal P}$ and ${\mathcal T}$ symmetry broken QWZ model in  section~\ref{Sec-IV}. In section~\ref{Sec-V}, we explore the nonreciprocity in the intraband and interband plasmon modes in ${\mathcal P}$ and ${\mathcal T}$ broken small angle TBG. Finally, we summarize our findings in section~\ref{Sec-VI}.    

\section{Theory of Intrinsic nonreciprocal intraband plasmons}\label{Sec-II}
In this section, we first explore the \textit{intrinsic} nonreciprocity in intraband plasmon dispersion for metallic systems. 
Plasmons are the isolated poles of the interacting density-density response function~\cite{giuliani2005quantum, Pines1962, fetter2003quantum, Sadhukhan_A_2017, Ghosh2017, Agarwal_L_2014}. 
The plasmon frequency can be calculated from the zeros of the dynamical dielectric function - $\varepsilon(\bm{q},\omega)$~\cite{giuliani2005quantum,fetter2003quantum,giuliani2005quantum, Thakur_D_2017, Sachdeva_P_2015}.
In random-phase approximation (RPA), the dielectric function is expressed as~\cite{giuliani2005quantum,Atasi_2022,Sadhukhan_N_2020,Dutta_C_2022} 
\be
\varepsilon(\bm{q},\omega)= 1 - V_{|\bm{q}|}^{(d)}\Pi(\bm{q},\omega)~.
\label{Eq_epsilon}
\ee
Here $V_{|\bm{q}|}^{(d)}$ denotes the Fourier transform of the Coulomb potential in $d$-dimension (for more information, see Appendix~\ref{appendix_Vq}).
We calculate the non-interacting density-density response function or the electron polarization function~\cite{giuliani2005quantum, Levitov2019} via, 
\bea
\Pi(\bm{q},\omega)= g\sum_{\bm{k}}\sum_{s,s^{\prime}}\frac{\left(f_{s,\bm{k}+\bm{q}} -f_{s^{\prime},\bm{k}} \right) F^{ss^{\prime}}_{\bm{k}+\bm{q},\bm{k}} }{E_{s,\bm{k}+\bm{q}}- E_{s^{\prime},\bm{k}}-\omega -i\eta}~.
\label{Pi_Eq1}
\eea
Here, $\sum_{\bm{k}}$ sums over the Brillouin zone (BZ), $s,~s^{\prime}$ are the band indices, {$E_{s,\bm{k}}$ denotes Bloch band energy with $|u_{s,\bm{k}}\rangle$ being the periodic part of Bloch function in the sth band}, $g$ denotes the degeneracy factor and $\eta$ is the broadening parameter. {We set $\hbar=1$ in our calculation.} The Fermi-Dirac distribution function for a given chemical potential $\mu$ is specified by $f_{s,\bm{k}}=\left[1+\exp\{(E_{s,\bm{k}}-\mu)/k_BT\} \right]^{-1}$. The coherence factor $F^{ss^{\prime}}_{\bm{k}+\bm{q},\bm{k}}= |\langle u_{s,\bm{k}+\bm{q}}|u_{{s}^{\prime},\bm{k}}\rangle|^2$ describes the overlap between two eigenstates of the Hamiltonian (${\cal H}_{\bm{k}}$) at momentum $\bm{k}$ and $\bm{k}+\bm{q}$, for band indices $s$ and $s^{\prime}$.

To examine the dependence of the plasmon dispersion on momentum transfer ${\bm q}$, we expand the polarization function in the dynamical long wavelength limit (${\bm q}\to 0$ and $\omega>q v_F $, where $v_F$ denotes the Fermi velocity) in powers of $1/\omega$ (see Appendix~\ref{sec_A1} for details)
\bea
\Pi_{\rm intra}(\bm{q},\omega)
&=& \frac{A_1(\bm{q})}{\omega} + \frac{A_2(\bm{q})}{\omega^2}+ \frac{A_3(\bm{q})}{\omega^3}+\dots~.
\label{Pi_intra}
\eea
Here, the expansion coefficients are calculated to be  
\be
A_{n}(\bm{q})=g\sum_{\bm{k}}{f_{s,\bm{k}}}\left[ F^{ss}_{\bm{k}+\bm{q},\bm{k}} \Delta E_{\bm{k}+\bm{q},\bm{k}}^{n-1} - F^{ss}_{\bm{k},\bm{k}-\bm{q}}\Delta E_{\bm{k},\bm{k}-\bm{q}}^{n-1}\right],
\label{Eq_A_n}
\ee
with $\Delta E_{\bm{k},\bm{k}^{\prime}}^{n}\equiv (E_{s,\bm{k}}-E_{s,\bm{k}^{\prime}})^n$ corresponding to $n$th power of the energy difference of the same band at different momentum. 
Here, the important quantity is the band coherence factor, $F^{ss}_{\bm{k}\pm \bm{q},\bm{k}}$. The intraband coherence term can be Taylor expanded up to ${\mathcal O}(q^3)$ as~\cite{Matsuura2010}
\be
F^{ss}_{\bm{k\pm q},\bm{k}}=1- q_aq_bg^{ab}_{s} \mp \frac{q_aq_bq_c}{2}\partial_{k_a}g^{bc}_{s} + {\cal O}(q^4)~,
\label{Eq_F}
\ee
and the repeated indices are summed. 
Here, $g^{ab}_{s}(\bm{k})$ represents intraband quantum metric (also known as  the
Fubini-Study metric)~\cite{Provost1980,Xiao_PRL2019}. 
For Bloch bands, the quantum metric $g^{ab}_{s}(\bm{k})$ defines the quantum distance between two infinitesimally close Bloch states at momentum $\bm{k}$ and $\bm{k}+d\bm{k}$~\cite{Resta2011}. {This gauge-invariant quantum metric is defined as~\cite{Matsuura2010} 
\bea
g^{ab}_{s}(\bm{k})=  {\rm Re}\left[\langle \partial_{k_a}u_{s,\bm{k}}|\partial_{k_b} u_{s,\bm{k}}\rangle\right] -{\cal R}^{a}_{ss}{\cal R}^{b}_{ss} ~,
\eea
where ${\cal R}^a_{ss}=i\langle u_{s,\bm{k}}|\partial_{k_a}u_{s,\bm{k}}\rangle$ is the intraband Berry connection.}

The power series expansion of Eq.~\eqref{Pi_intra}, specifically the odd power of $1/\omega$, is crucial for describing the nonreciprocal plasmons.  
By approximating the polarization function, $\Pi(\bm{q},\omega)$ up to $1/\omega^3$ order of terms, we can approximately evaluate the plasmon dispersion by solving for the roots of $\varepsilon(\bm{q},\omega)$ in Eq.~\eqref{Eq_epsilon}. This leads  to a cubic equation
\bea
\omega^3- V_{|\bm q|}^{(d)}\left[A_1(\bm{q})\omega^2+ A_2(\bm{q})\omega + A_3(\bm{q}) \right]=0~.
\label{Eq_roots}
\eea
The solutions to this cubic equation aid in understanding the origin of nonreciprocity in the plasmon dispersion. 
We first investigate the long-wavelength expansion of each of these expansion coefficients.
The first expansion coefficient $A_1(\bm{q})$ can be calculated up to the third order of $q$ as 
{
\bea
A_{1}(\bm{q})&=&q_aq_bq_c{\cal Q}_{abc} + {\mathcal O}(q^4)~.
\label{Eq_A1}
\eea
Here, ${\cal Q}_{abc}=- g\sum_{\bm{k}}f_{s,\bm{k}}\partial_{k_a}g^{bc}_{s}(\bm{k})$ represents the quantum metric dipole; the dipole moment of the quantum metric of occupied states in the momentum space. This is analogous to the Berry-curvature dipole, which is the origin of quantum nonlinear Hall effect~\cite{Sodemann2015,atasi2022N}.}
Note that in presence of ${\mathcal T}$ or ${\mathcal P}$ symmetry, ${\mathcal Q}_{abc} \to 0$ as shown in Appendix~\ref{symmetry_G_abc}. 

Similarly, we can expand the second coefficient in the small $q$ limit to obtain, 

{
\bea
A_2(\bm{q})&=& q_aq_b{\cal D}_{ab}+{\mathcal O}(q^4)~.
\label{Eq_A2}
\eea
Here, ${\cal D}_{ab}=\left[-g\sum_{\bm{k}}\frac{\partial f_{s,\bm{k}} }{\partial E_{s,\bm{k}}}v_{s,\bm{k}}^{a}v_{s,\bm{k}}^{b} \right]$ represents the Drude weight, and $v_{s,\bm{k}}=\nabla_{\bm{k}}E_{s,\bm{k}}$ is the band velocity.} The Drude weight~\cite{Atasi_2022,Agarwal2011} is completely a Fermi-surface property. In the presence of a finite Fermi surface, $A_2(\bm{q})$ is non-zero regardless of whether the ${\cal P}$ or ${\cal T}$ symmetry is present or absent. {Interestingly, $A_2(\bm{q})$ does not contain any odd power of $q$ expansion coefficients.} 
The third term of Eq.~\eqref{Pi_intra} can be calculated as
\bea
A_3(\bm{q})&=& q_aq_bq_c{\cal C}_{abc}   + {\mathcal O}(q^4)~.
\label{A3}
\eea
{Here, ${\cal C}_{abc}=\left[-2g \sum_{\bm{k}}\frac{\partial f_{s,\bm{k}} }{\partial E_{s,\bm{k}}}v_{s,\bm{k}}^{a}v_{s,\bm{k}}^{b}v_{s,\bm{k}}^{c}\right]$ denotes the higher order Drude-weight, arising due to the asymmetry of the electron band velocity over the Fermi surface.} In absence of ${\mathcal P}$ and ${\mathcal T}$ symmetries, ${\mathcal C}_{abc}$ becomes finite, as shown in Appendix~\ref{symmetry_C_abc}.

In the presence of ${\mathcal P}$ or ${\mathcal T}$ symmetry in the system, the odd $1/\omega$ power expansion terms of $\Pi(\bm{q},\omega)$ in Eq~\eqref{Pi_intra} vanish and only even powers survive. 
This leads to reciprocal plasmon dispersion $\omega_p^{0}(\bm{q})=\sqrt{V_{|\bm q|}^{(d)}q_aq_b{\mathcal D}_{ab}}$ on account of $A_2(\bm{q})$ being an even function of ${\bm q}$.
In contrast to this, when both the ${\mathcal P}$, and ${\mathcal T}$ symmetries are simultaneously broken, the odd $1/\omega$ power expansion coefficients, such as $A_1(\bm{q})$, $A_3(\bm{q})$ become finite and contribute to the plasmon dispersion. The exact plasmon dispersion can be solved from Eq.~\eqref{Eq_roots}. However, generally we have $A_2> (A_1$, $A_3)$, and we can perturbatively solve Eq.~\eqref{Eq_roots} up to first order of $A_3$ and $A_2$ in (see Appendix~\ref{Eq.12_derivation} for detailed derivation)
\bea
&&\omega_p^{\rm intra}(\bm{q})\approx \omega_p^{0} + \frac{1}{2}V_{|\bm{q}|}^{(d)}A_1  +\frac{A_3}{2A_2} + {\mathcal O}\left(A_1,A_3\right)^2~,\nn
\\
&&\approx \sqrt{V_{|{\bm q}|}^{(d)}{q_aq_b{\mathcal D}_{ab}} } +  \frac{1}{2}V_{|\bm{q}|}^{(d)}q_aq_bq_c{\mathcal Q}_{abc} + \frac{q_aq_bq_c}{q_mq_n}\frac{{\mathcal C}_{abc}}{2{\mathcal D}_{mn}}~.\nn
\\
\label{Plasmon_disp}
\eea
Interestingly, for any finite values of ${\mathcal Q}_{abc}$ and ${\mathcal C}_{abc}$,  the plasmon eigenmode of Eq.~\eqref{Plasmon_disp} is an asymmetric 
function of the wave propagation direction $\bm{q}$, irrespective of the dimensions. Hence, the second and third terms of Eq~\eqref{Plasmon_disp} are the new sources of intrinsic plasmonic nonreciprocity in ${\mathcal P}$ and ${\mathcal T}$ broken quantum systems. The second term of Eq.~\eqref{Plasmon_disp} is dictated by the quantum metric dipole, originating from the asymmetry of the quantum metric over the Fermi surface. This term depends on the geometric properties of the electronic wave function. The contribution of nonreciprocity through the $A_3(\bm{q})$ term is independent of the quantum metric and is governed by the asymmetric velocities of the electron wave packet over the Fermi surface in different directions. 

Our analysis establishes that the simultaneous breaking of both ${\cal P}$ and ${\cal T}$ symmetry in the system gives rise to nonreciprocity in the intraband plasmon dispersion. This will manifest in noncentrosymmetric magnetic materials~\cite{ShouChengZhang2016, BGhosh2021}.  
The plasmon modes appear as peaks in the energy loss function spectrum, ${ L}({\bm q},\omega)$, which measures the amount of energy a system can absorb from an external perturbation with wavevector $\bm{q}$ and energy $\hbar\omega$.
The energy loss function is related to the dielectric function via~\cite{EELS2014}
\begin{equation}
{L}(\bm{q},\omega) \approx  -{\rm Im}\left[\frac{1}{\varepsilon(\bm{q},\omega)}\right].
\label{EELS}
\end{equation}
From an experimental point of view, we can measure the difference in the loss function spectra; $\delta L(q,\omega)\equiv |L(-\bm{q},\omega)-L(\bm{q},\omega)|$. This will capture two distinct plasmon peaks at different energies for the same $q$ value, in ${\mathcal P}$ and ${\mathcal T}$ broken materials~\cite{EELS2014}, highlighting the plasmonic nonreciprocity.

Below, we discuss the nonreciprocity in interband plasmons in the next section. Following that, we explicitly demonstrate the nonreciprocity in both the interband and the intraband plasmon mode in two different systems in Sec.~\ref{Sec-IV} and Sec.~\ref{Sec-V}, respectively. 

\section{Intrinsic nonreciprocity in interband plasmons}\label{section-III}
In this section, we study \textit{intrinsic} nonreciprocity in interband plasmons for multi-band systems. {The interband plasmon arises when the nested bands in the Brillouin zone give rise to a large number of nearly parallel interband electron-hole transitions at constant energy~\cite{Atasi_2022, FrankHLKoppen2021, Stauber2016}. 
This effect manifests as a prominent peak in the imaginary component of the dielectric function, accompanied by a corresponding root in the real part through Krammer's Kroning transformation~\cite{giuliani2005quantum}. The interband plasmon is gapped in all dimensions at $q=0$, a defining feature that stems from the singularity of the joint-density-of-states spectrum
\be
{\rm JDOS}(\omega)=\sum_{\bm{k},s,s^{\prime}}^{s^{\prime}\ne s}f_{s,\bm{k}}(1-f_{s^{\prime},\bm{k}})\delta(\omega+E_{s,\bm{k}}-E_{s^{\prime},\bm{k}})~.
\ee
} 
To investigate interband plasmon dispersion, we start from the interband polarization function in Eq.~\eqref{Pi_Eq1}, considering only interband transitions. 
\normalsize
\begin{figure*}[!t]
	\includegraphics[width=0.94\linewidth]{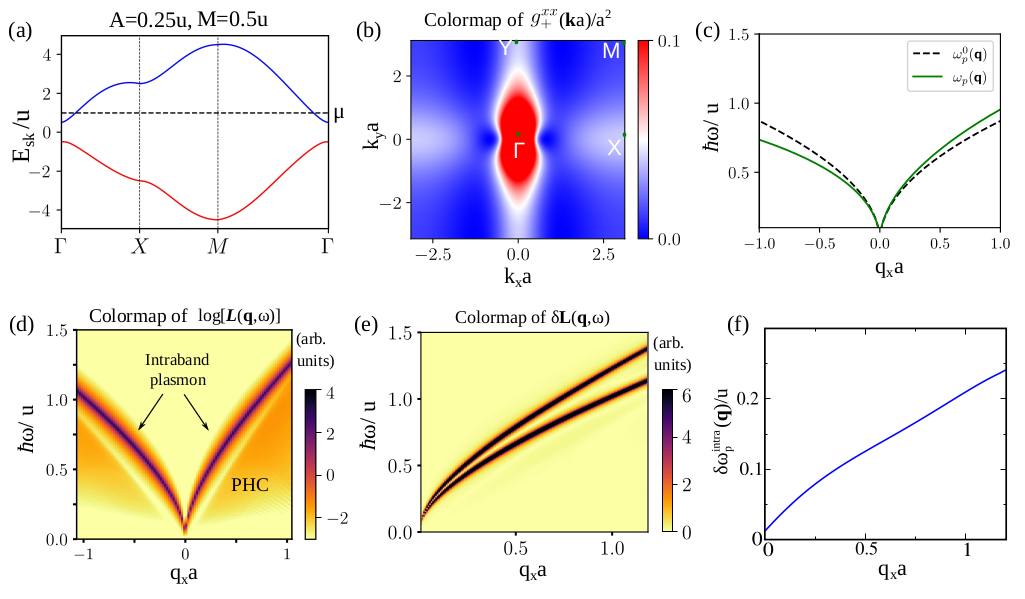}
	\caption{
	   (a) Electronic band dispersion of the 2D Qi-Wu-Zhang model [see Eq.~\eqref{QWZ_model_Eq}] with the parameters set $A$=$0.25u$ and $M$=$0.5u$. The QWZ Hamiltonian breaks both the parity and the time-reversal symmetry of the Hamiltonian. We set the chemical potential to be $\mu=1.0u$. (b) The variation of the quantum metric for the conduction band [$g^{xx}_{+}(\bm{k})$] over the Brillouin zone. It has a large value near the band edge around the $\Gamma$ point. (c) Nonreciprocal intraband plasmon dispersion (solid green line) for this system is obtained by solving for the roots of the dielectric function in Eq.~\eqref{Eq_roots}. The black dashed line represents reciprocal plasmon dispersion with the nonreciprocal terms switched off [$A_1$ and $A_3$ terms set to zero in Eq.~\eqref{Pi_intra}].  
	(d) The colormap of the energy loss function $L(\bm{q},\omega)$, calculated numerically from the interacting polarization function (within the random phase approximation). The nonreciprocity of the plasmon dispersion can be clearly seen.
    (e) The difference in the loss function $\delta L({\bm q},\omega)=|L({\bm q},\omega)-L(-{\bm q},\omega)|$ for equal but opposite momentum transfer captures the nonreciprocity via two distinct plasmon energies for each momentum. (f) Numerically calculated asymmetry of the intraband plasmon dispersion, $\delta\omega_p^{\rm intra}(\bm{q})$ as a function of wavevector. 
    \label{QWZ_fig_intra}}
\end{figure*}
%
The interband coherence term is Taylor expanded in different powers of $q$ as (see Appendix~\ref{appendix_H})
\bea
F^{ss^{\prime}}_{\bm{k},\bm{k}\pm \bm{q}}=q_aq_bF^{(2)}_{ab,ss^{\prime}} \pm q_aq_bq_c F^{(3)}_{abc, ss^{\prime}}+\mathcal{O}(q^4)~,
\label{Interband_overlap_Eq}
\eea
{where, $a$, $b$, $c$ denotes Cartesian directions and $s$, $s^{\prime}$ represents band indices.}
{Here, the expansion coefficients are calculated as,
\bea
F^{(2)}_{ab,ss^{\prime}}&=&{\cal R}^{a}_{ss^{\prime}}{\cal R}^{b}_{s^{\prime}s}~,\nn
\\
F^{(3)}_{abc,ss^{\prime}}&=&{\rm Re}\left[ \langle u_{s,\bm{k}} | \partial_{{k}_a} u_{s^{\prime},\bm{k}}\rangle \langle \partial_{{k}_b}\partial_{{k}_c}u_{s^{\prime},\bm{k}}| u_{s,\bm{k}}\rangle \right]~,
\eea
where ${\cal R}^a_{ss^{\prime}}$($\equiv i\langle u_{s,\bm{k}}|\partial_{\bm{k}_a}u_{s^{\prime},\bm{k}}\rangle$) represents band resolved Berry connection.} For $a$=$b$=$c$, $F^{(2)}_{ab,ss^{\prime}}$ represents square modulus of the Berry connection, and $F^{(3)}_{abc,ss^{\prime}}$ equals to the metric connection ($\Gamma_{s^{\prime}s}^{abc}$)~\cite{Agarwal2022A,AhnPRX2020},-- a quantum {geometric quantity defined in Eq.~\eqref{Eq_L1} of Appendix~\ref{Gamma_mp}}.

For an approximate but insightful estimation of the interband plasmon dispersion, we consider a pair of nested or nearly flat bands over BZ and approximate $E_{s^{\prime},\bm{k}+\bm{q}}- E_{s,\bm{k}}\approx \Delta_0$, where $\Delta_0$ is the peak in JDOS~\cite{Atasi_2022}. {This leads to
(see Appendix~\ref{appendix_interband_Pi} for more details), 
\be
\Pi_{\rm inter}(\bm{q},\omega)\approx \frac{2}{(\omega^2-\Delta_0^2)}\left[q_aq_b{\cal A}_{ab}\Delta_0 + q_aq_bq_c{\cal B}_{abc}\omega\right],
\label{Pi_inter_expansion}
\ee
where the different band-geometric contributions are expressed as,
\be
{\cal A}_{ab}= g\sum_{\bm{k}}\sum_{s, s^{\prime}}^{s^{\prime}\ne s} f_{s,\bm{k}}{\cal R}^{a}_{ss^{\prime}}{\cal R}^{b}_{s^{\prime}s}~,
\label{def_A}
\ee
\be
{\cal B}_{abc}= g\sum_{\bm{k}}\sum_{s, s^{\prime}}^{s^{\prime}\ne s} f_{s,\bm{k}} F^{(3)}_{abc,ss^{\prime}}~.
\label{def_B}
\ee
Here, ${\cal A}_{ab}$ is generally finite, independent of the presence or absence of either ${\cal P}$ or ${\cal T}$. But, ${\cal B}_{abc}$ vanishes in presence of either ${\cal P}$ or ${\cal T}$ symmetries (see Appendix~\ref{symmetry_B_abc} and Appendix~\ref{symmetry_A} for details).}

Now, the interband plasmons are simply the zeros of the real part of the dielectric function, which yields
\bea
1-V_{|\bm{q}|}\Pi_{\rm inter}(\bm{q},\omega)=0~.
\label{epsilon_inter}
\eea
We choose the constant wavevector $\bm{q}=q\hat{\bm{x}}$, and define ${\cal A}=g\sum_{\bm{k}}\sum_{(s\ne s^{\prime})}f_{s,\bm{k}}|{\cal R}_{ss^{\prime}}^{x}|^2$, and ${\mathcal B}=g\sum_{\bm{k}}\sum_{(s\ne s^{\prime})}f_{s,\bm{k}}{\Gamma}_{s^{\prime}s}^{xxx}$, respectively.
Now, by solving for the roots of Eq.~\eqref{epsilon_inter}, we obtain the interband plasmon dispersion
\bea
\omega^{\rm inter}_{p}(\bm{q}) \approx q^3V^{(d)}_{|\bm{q}|}{\mathcal B} &+& \left[ \left(q^3V^{(d)}_{|\bm{q}|}{\mathcal B}\right)^2 + \Delta_0^2
\right. \nn
\\
&&
\left.
 +2V^{(d)}_{|\bm{q}|}\Delta_0{\mathcal A}q^2 \right]^{1/2}~.
\label{omega_inter}
\eea
Owing to the $q^3$ term with ${\cal B}$, Equation~\eqref{omega_inter} gives us a nonreciprocal interband plasmon dispersion which is asymmetric for $+\bm{
q}$ and $-{\bm q}$ wavevector. So, in the absence of both ${\cal P}$ and ${\cal T}$ symmetry, the interband plasmon dispersion becomes {\it intrinsically} nonreciprocal due to the band geometric quantity, ${\cal B}$. 
But, in the presence of either ${\cal P}$ or ${\cal T}$ symmetry, ${\cal B}= 0$, and Eq.~\eqref{omega_inter} reduces to the reciprocal interband plasmon dispersion~\cite{Atasi_2022},  
\be
\omega^{\rm inter}_{0}(\bm{q})\approx \Delta_0\sqrt{1+\frac{ 2V^{(d)}_{|\bm{q}|}{\mathcal A}}{\Delta_0}q^2 }~.
\ee 
This establishes that similar to the case of the intraband plasmon, the nonreciprocity in the interband plasmon also manifests in non-centrosymmetric magnetic systems due to the quantum geometry of Bloch states.

Next, we investigate this nonreciprocity in the plasmon dispersion in two different systems in section~\ref{Sec-IV} and section~\ref{Sec-V}. 

\section{Plasmonic Non-reciprocity in ${\mathcal P}$ and ${\mathcal T}$ broken 2D Qi-Wu-Zhang Model}\label{Sec-IV} 
{In this section, we study nonreciprocal plasmons for the `Qi-Wu-Zhang' model Hamiltonian~\cite{QiWuZhang2006, Asb_th_2016,Hughes2019}}. It is specified on a 2D square lattice as 
{
\bea
{\cal H}_{\bm{k}}&=&A\sin({k_x}a)\mathbb{I} + [M+2u-u\cos({k_x}a)-u\cos({k_y}a)]\sigma_z \nn
\\
& &+u\sin({k_x}a)\sigma_x + u\sin({k_y}a)\sigma_y~.
\label{QWZ_model_Eq}
\eea}%
For $A=0$, and $-4<M<-2$ or $-2<M<0$, this model had been proposed to describe a chern insulator phase~\cite{Asb_th_2016}.
For this fermionic system, the parity (${\mathcal P}$) and time reversal (${\mathcal T}$) operator have a definite representation given by, ${\mathcal P}  \leftrightarrow \sigma_z$ and ${\mathcal T} \leftrightarrow \bm{K}$, where $\bm{K}$ is the anti-Hermitian complex conjugation operator~\cite{Wanatabe_PhysRevX2021}. This model intrinsically breaks time-reversal symmetry, i.e; ${\mathcal T} {\cal H}_{\bm{k}}{\mathcal T}^{-1}\ne {\cal H}_{-\bm{k}}$. For $A=0$, this model preserves inversion symmetry, i.e; $\sigma_z {\cal H}_{\bm{k}} \sigma_z^{-1}={\cal H}_{-\bm{k}}$, while for 
$ A \neq 0$, it does not. Below, we discuss the nature of intraband and interband plasmon dispersion for the QWZ model in the absence of both ${\mathcal P}$ and  ${\mathcal T}$ symmetry. 
\subsection{Intraband plasmon}

The band dispersion of the two-band QWZ model is shown in Fig.~\ref{QWZ_fig_intra} (a), for the choice of $A=0.25 u$ and $M=0.5 u$. The upper band has a minimum at $\Gamma$ and a maxima at the BZ corner  $\bm{k}a=(\pm \pi,\pm \pi)$. The distribution of the quantum metric $g^{xx}_{+}(\bm{k})$ for the conduction band over BZ is shown in Fig.~\ref{QWZ_fig_intra}(b). The quantum metric shows a peak near the band edge at $\Gamma$ point, which is the hotspot of different geometrical quantities~\cite{kamaldas2022_arxiv}. 

To demonstrate intraband plasmonic nonreciprocity, we compute the various expansion coefficients $A_1(\bm{q})$, $A_2(\bm{q})$ and $A_3(\bm{q})$ numerically using  Eq.~\eqref{Pi_intra}. 
The plasmon dispersion is then calculated by solving for the zeros of the dielectric function $\varepsilon(\bm{q},\omega)$. We choose $\bm{q}=(q_x,0)$ and the dimensionless parameter $e^2/(\kappa\epsilon_0ua)=50$ for  our numerical calculation. In Fig.~\ref{QWZ_fig_intra}(c), we present the nonreciprocal plasmon dispersion by solving Eq.~\eqref{Eq_roots}. The intrinsic nonreciprocity is dominated by the combined effect of the quantum metric dipole (${\mathcal Q}_{abc}$) and the higher order Drude weight(${\mathcal C}_{abc}$). 
This treatment is approximate and valid for a small ${\bm{q}}$ limit compared to the Fermi wavevector. Going beyond the small ${\bm q}$ limit, we also compute the exact polarization function $\Pi(\bm{q},\omega)$ by using Eq.~\eqref{Pi_Eq1} and the RPA dielectric function $\varepsilon(\bm{q},\omega)$. 
In Fig.~\ref{QWZ_fig_intra}(d), we have shown the colormap of the loss function spectrum in the $\bm{q}$-$\omega$ plane. The nonreciprocity of the bulk plasmon spectrum, which lies outside the particle-hole continuum (PHC) region~\cite{giuliani2005quantum}, can be clearly seen in Fig.~\ref{QWZ_fig_intra}(d). In Fig.~\ref{QWZ_fig_intra}(e), we present the difference in the loss function spectrum in the $\bm{q}$-$\omega$ plane. This shows two peaks in loss function for the same magnitude of wavevector but pointing in opposite directions. The asymmetry of the dispersion; $\delta\omega_p^{\rm intra}(\bm{q})$ monotonically increases with wave vector as shown in Fig.~\ref{QWZ_fig_intra}(f).

\subsection{Interband plasmon}
\begin{figure}[h!]
    \includegraphics[width=\linewidth]{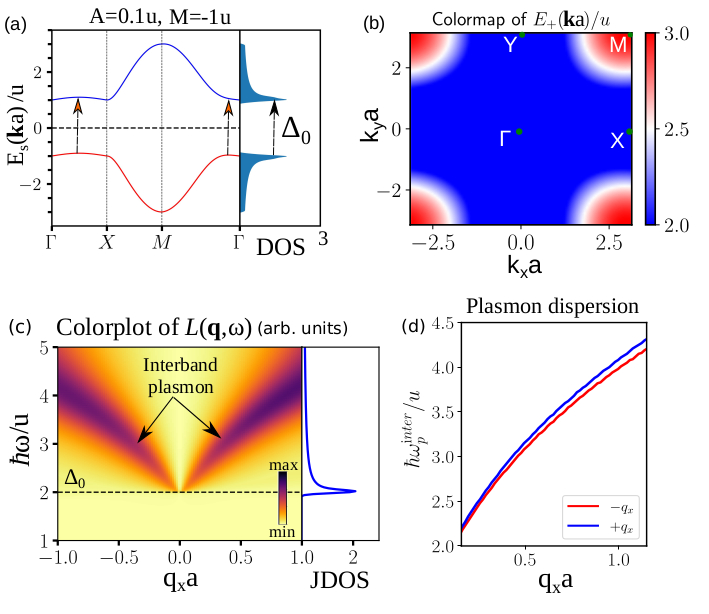}
    \caption{ (a) Electronic band dispersion and density of states of the Qi-Wu-Zhang model [see Eq.~\eqref{QWZ_model_Eq}] with $A=0.1$u and $M=-1.0$u. With these parameters, the QWZ model breaks both parity and time-reversal symmetry. (b) Colormap of the conduction band, $E_{+}(\bm{k})$ over Brillouin zone, which captures constant energy dispersion. (c) Colormap of $L(\bm{q},\omega)$ captures the gapped interband plasmon. {The plasmon gap ($\Delta_0$) is specified by the peak of the joint density of states spectrum.} (d) To highlight the small nonreciprocity in the interband plasmon dispersion, we have numerically solved for the plasmon poles of the dielectric function $\varepsilon(\pm\bm{q},\omega_p)=0$. The interband plasmon dispersion in (d) clearly shows the small but finite difference in the plasmon energy eigenmode for $\pm\bm{q}$ wave vectors. 
    }
    \label{QWZ_fig_interband}
\end{figure}

To demonstrate interband plasmon and intrinsic nonreciprocity, we choose a ${\mathcal P}$ and ${\mathcal T}$ broken QWZ model with $A=0.1u$ and $M=-1.0u$ as input parameters in Eq.~\eqref{QWZ_model_Eq}. 
We present the corresponding electronic band dispersion in Fig.~\ref{QWZ_fig_interband}(a).
The conduction and valence bands are almost flat along the $\Gamma-X$ path,  giving rise to VHS peaks in the DOS spectrum. Both the conduction and valence bands are dispersive around $M$, as shown in the colormap of eigenvalues of the conduction band ($E_{+}(\bm{k})$) over BZ in Fig.~\ref{QWZ_fig_interband}(b). 

{The nearly flat valence and conduction bands induce nearly parallel electron-hole transitions in momentum space, resulting in a singularity in the JDOS spectrum. This effectively gives rise to an interband plasmon mode, as shown in the colorplot of $L(\bm{q},\omega)$ in Fig.~\ref{QWZ_fig_interband}(c).}
The small nonreciprocity in the interband plasmon dispersion is not visibly clear from {loss function spectrum}. To resolve this, we present the numerically calculated plasmon dispersion 
for $+\bm{q}$ and $-\bm{q}$ in Fig.~\ref{QWZ_fig_interband}(d). This shows distinct interband plasmon poles in $\omega$- plane for the same magnitude of wavevector ($|\bm{q}|$) with opposite directions. The origin of the interband nonreciprocity is associated with the non-zero value of ${\cal B}$ [see Eq.~\eqref{def_B}] in the absence of both ${\cal P}$ and ${\cal T}$ symmetries. In the presence of either of these symmetries, ${\cal B}$ turns out to be zero and leads to reciprocal interband plasmon as described in Sec.~\ref{section-III}. This has been illustrated in Appendix~\ref{Appendix_reciprocal_interband}.            
This highlights the presence of \textit{intrinsic} nonreciprocal interband plasmon in ${\mathcal P}$ and ${\mathcal T}$ broken systems. 
We present another example of interband nonreciprocity in a one-dimensional magnetic bipartite lattice model in Appendix~\ref{Appendix-K}. 

Below, we discuss the possibility of \textit{intrinsic} nonreciprocal intraband and interband plasmon modes in realistic moir\'{e} superlattices of twisted bilayer graphene.
\begin{figure*}[t!]
	\includegraphics[width=0.9\linewidth]{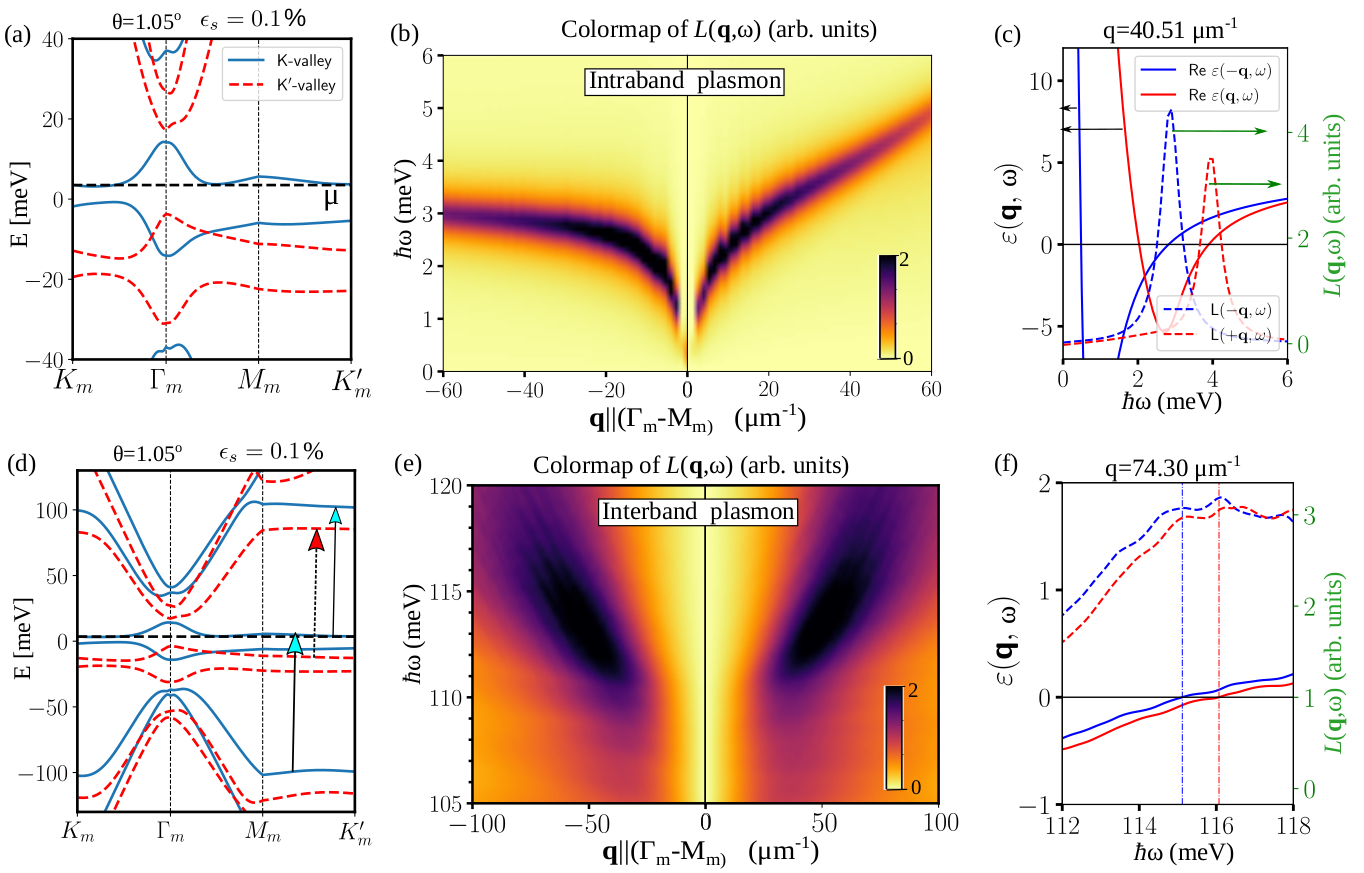}
	\caption{  
    (a) Band dispersion of 
    \textit{magic} angle ($\theta=1.05^{\circ}$) twisted bilayer graphene with strain ($\epsilon_s=0.1{\%}$) and in the presence of a gap of 17 meV. The spontaneous time-reversal symmetry breaking is introduced via valley polarization by shifting the $K^{\prime}$ valley by $\Delta_{vs}=15$ meV.
    (b) The energy loss function $L(\pm\bm{q},\omega)$ for $\bm{q}$ along $\Gamma_{m}$-$M_{m}$ clearly showing the nonreciprocity in the intraband plasmon propagation for $-{\bm{q}}$ and $+{\bm q}$ direction. The chemical potential is set to  $\mu$=$3.5$ meV at $K$-valley, and we work at zero temperature.
    This nonreciprocity is also captured by the distinct zeros of the real part of the dielectric function in (c). (d) Band dispersion of TBG in higher energy window. 
    {The interband plasmon arises due to the vertical electron-hole transitions from flat bands to higher moir\'{e} minibands as marked by arrows in panel (d). }
    (e) The nonreciprocal interband plasmons can be seen in the color plot of the loss function or in the distinct roots of the dielectric function for $\pm {\bm q}$ (marked by vertical lines) in panel (f). 
    }
    \label{TBG_fig}
\end{figure*}

\section{Intrinsic nonreciprocal plasmons in twisted bilayer graphene}\label{Sec-V}
In quest of a more realistic example,  we explore the existence of nonreciprocal plasmon modes in moir\'{e} systems~\cite{Santos2007G,Bistritzer2011,Morell2017T, Koshino2018M,Stauber2018C,Stauber2018L, Sinha_B_2022, atasi2022N, Atasi_2022, Sinha2020, Adak2020}. Twisted graphene heterostructures are fabricated by stacking multiple free-standing graphene monolayers on top of each other with small relative rotations. These moir\'{e} systems generally host slow (dispersionless) and highly tunable plasmon modes due to substantial interband transition between the nested sub-bands~\cite{Stauber2016, Levitov2019, Atasi_2022}. Motivated by the recent experimental verification of the theoretically proposed novel interband plasmon mode of TBG through mid-infrared near-field optical microscopy~\cite{FrankHLKoppen2021}, we specifically focus on the nature of plasmon modes for \textit{magic} angle TBG. 
These novel plasmon modes can enable strong light-matter interactions within the highly sought-after mid-wave infrared spectral range~\cite{Huang2022, Bahram2006P}. In this section, we explore {\textit intrinsically} nonreciprocal interband and intraband plasmon modes in TBG with broken ${\cal P}$ and ${\cal T}$. 

{To construct the low energy model of TBG system, we have adapted the continuum model Hamiltonian approach (see Appendix~\ref{Appendix_continuum_model} and Appendix~\ref{Matrix_strain}) originally proposed by Santos-Peres-Neto~\cite{Santos2007G} and later also reported by Bristizer and Macdonald~\cite{Bistritzer2011}. In reality, the pristine twisted bilayer graphene exhibit structural chirality and intrinsically breaks inversion symmetry~\cite{Morell2017T,Stauber2018C,Stauber2018L}. However, the adapted valley-specific low energy continuum model Hamiltonian with enforced spin degeneracy fails to capture the information of the overall inversion symmetry breaking of the  system~\cite{pantaleon2021T,Chaudhary2022S}. So, to mimic the breaking of inversion symmetry, we use a staggered sublattice potential term, similar to that generated by the hBN substrate on the graphene layer, which is known to break the inversion in single-layer graphene~\cite{Wang2016G}. }

{To satisfy the second criterion of breaking time-reversal symmetry to achieve nonreciprocity of plasmons, we have introduced a valley asymmetry between the $K$ and $K^\prime$ valleys. This lifting of valley degeneracy, sometimes referred to as orbital ferromagnetism, is known to break the effective time-reversal symmetry of the system~\cite{Liu2020}. The valley splitting can arise due to either an externally applied weak magnetic field or intrinsic strong electron-electron interaction induced spontaneous time-reversal symmetry breaking in twisted graphene multilayers~\cite{Liu2019Q,Sharpe2019E}. } 

{To investigate the nature of plasmonic dispersion, we choose h-BN aligned TBG with magic angle twist ($\theta \approx 1.05^\circ$). The h-BN substrate, essential to fabricate TBG devices, induces finite strain in TBG. To capture the realistic experimental scenario, we have introduced a uniaxial strain ($\epsilon_s$) along the zigzag direction of TBG (see Appendix~\ref{Matrix_strain} for details). This interfacial strain breaks all the spatial symmetries except the $C_{2z}$ rotational symmetry [see Fig.~\ref{moire_BZ} (c)]. The band-dispersions of $K$ and $K^\prime$ valley, including the effect of the sublattice symmetry breaking ($\Delta=$17 meV), uniaxial strain ($\epsilon_s=0.1\%$), and finite valley splitting of $\sim$15 meV, are shown in Fig.~\ref{TBG_fig}(a) and (d).} The flat bands near the Fermi energy give rise to VHS in the DOS. Utilizing the obtained energy eigenvalues and eigenfunctions, we numerically calculate the RPA dielectric function and energy loss function spectrum, $L(\bm{q},\omega)$ including all intra ($s=s^{\prime}$) and interband ($s\ne s^{\prime}$) transitions. The 2D coulomb potential of the form $V^{(2)}_{\bm{q}}=2\pi e^2/{\kappa |\bm{q}|}$ is used in our calculations where $\kappa=3.03$ being the static background dielectric constant for hBN/TBG/air interface~\cite{Levitov2019}.

The color plot of the loss function distribution $L({\bm q},\omega)$, arising from the low energy intraband contribution, is shown in Fig.~\ref{TBG_fig}(b) for both positive and negative momentum transfer parallel to $\Gamma_m-M_m$ high symmetry direction. The asymmetry of the plasmon dispersion between the $+\bm{q}$ and $-\bm{q}$ clearly suggests the clear signature of nonreciprocity. For visual clarity, we also show the line cut of $L(\bf{q},\omega)$ for fixed $q=$40.51$\mu m^{-1}$ in Fig.~\ref{TBG_fig}(c). The nonreciprocity is further substantiated by the distinct zeros of the real part of the dielectric function, $\varepsilon(\bm{q},\omega)$ for $+{\bm q}$ and $-{\bm q}$ wavevectors (see Fig.~\ref{TBG_fig}(c)). {Moreover, we have found that while the substrate-induced strain can increase the degree of nonreciprocity of plasmons, it is not a necessary criterion to observe this phenomenon. In Appendix~\ref{Appendix_continuum_model}, we show that ${\cal P}$ and ${\cal T}$ broken TBG also exhibits small but finite nonreciprocity in plasmon dispersion even without strain (see Fig.~\ref{strain_intra}).}

In addition to the intraband plasmon, TBG also supports interband plasmon mode due to the collective motion of interband transitions from flat band to moir\'{e} minibands. In Fig.~\ref{TBG_fig}(d), we show the band-dispersion within a broad energy range marking the interband electronic transitions by vertical arrows. The energy loss function corresponding to the interband plasmon is shown in Fig.~\ref{TBG_fig}(e). 
Due to the absence of both ${\cal P}$ and ${\cal T}$ symmetry, this propagating interband plasmon mode also shows nonreciprocal nature. The magnitude of nonreciprocity of interband modes is relatively smaller (though finite), due to smaller interband coherence. The existence of this nonreciprocity on the propagation direction is easily visualized from the line cuts of the dielectric function and energy loss function of Fig.~\ref{TBG_fig}(f) for $q=$74.3 $\mu m^{-1}$. 

Our calculations predict \textit{intrinsic} nonreciprocal intraband and interband plasmon modes in TBG arising from the band geometric terms. This nonreciprocal dynamics of the bulk plasmon opens a new avenue to explore direction-specific magneto-chiral optical effects in moir\'{e} superlattices.

\section{Conclusion}\label{Sec-VI}
Nonreciprocity in plasmon propagation manifests as the different frequencies of the plasmon modes propagating in opposite directions. 
In this work, we present an analytical formulation for understanding \textit{intrinsic}  nonreciprocal bulk plasmon modes depending on the underlying symmetries of the system. 
We highlight the role of band geometric quantities in generating {\it intrinsic} bulk plasmonic nonreciprocity even without external bias. We show that in the absence of ${\cal P}$ and ${\cal T}$  symmetry, the interband plasmon mode, as well as the intraband plasmon mode, will be \textit{intrinsically} nonreciprocal. For the interband plasmon mode, the $\bm{k}$- integral of the Fermi distribution function weighted metric connection [see Eq.~\eqref{omega_inter}] is the primary quantity which dictates the nonreciprocity. For the intraband plasmon modes, the quantum metric dipole (${\mathcal Q}_{abc}$) and higher order Drude weight-like term  (${\cal C}_{abc}$) are responsible for the nonreciprocity.
We explicitly demonstrate {\it intrinsic} intraband and interband nonreciprocity in the two-dimensional `Qi-Wu-Zhang'  model and twisted bilayer graphene moir\'{e} systems in the absence of both ${\cal P}$  and ${\cal T}$ symmetries. 
We hope our theoretical study will stimulate further investigations on unidirectionally propagating plasmons in noncentrosymmetric magnetic systems and pave the way for designing novel photonic devices.

\section{Acknowledgment}
\noindent  A.~C. acknowledges the Indian Institute of Technology, Kanpur and Science and Engineering Research Board (SERB) National Postdoctoral Fellowship (PDF/2021/000346), India for financial support. A.~A. acknowledges the Science and Engineering Research Board for Project No. MTR/2019/001520, and the Department of Science and Technology for Project No. DST/NM/TUE/QM-6/2019(G)-IIT Kanpur, of the Government of India, for financial support.
We thank Kamal Das, Debottam Mandal, and Sunit Das for the useful discussions. {We ac-
knowledge the high-performance computing facility at
IIT Kanpur for computational support. We also acknowl-
edge the National Supercomputing Mission (NSM) for providing computing resources of ‘PARAM Sanganak’
at IIT Kanpur.}
\appendix
\section{Coulomb potential in d-dimension}\label{appendix_Vq}
The Fourier transform of the Coulomb interaction $v(r)=e^2/(4\pi\kappa\epsilon_0r)$, in  $d$-dimension is given by
\bea
V_{|\bm{q}|}^{(d)}&=&\frac{4\pi e^2}{4\pi\kappa\epsilon_0q^2},~~d=3~,\nn
\\
&=& \frac{2\pi e^2}{4\pi\kappa\epsilon_0q},~~d=2~,\nn
\\
&=&-\frac{e^2}{4\pi\kappa\epsilon_0}e^{q^2R^2}Ei(-q^2R^2),~~d=1~.
\label{Eq_Vq}
\eea
Here, $\kappa$ is the background-material-dependent static dielectric
constant, $\epsilon_0$ denotes free space permittivity, and $Ei(x)$ is the exponential-integral function~\cite{giuliani2005quantum}. In one dimension, $R$ represents the characteristic of the lateral confinement size (say, the radius of 1D nanoribbon).

\section{Expansion of polarization function in small ${\bf q}$ limit}\label{sec_A1}
We start with the definition of density-density response or polarization function in Eq.~\eqref{Pi_Eq1}. To proceed further analytically, we first rewrite Eq.~\eqref{Pi_Eq1} by performing a standard replacement $\bm{k}\to \bm{k}-\bm{q}$ and $s\to s^{\prime}$ to the first term containing $f_{s,\bm{k}+\bm{q}}$~\cite{Lewandowski2020}. Then, the total response function can be divided into two parts depending on the directionality of momentum transfer $\bm{q}$ (parallel or anti-parallel to the wavevector $\bm{k}$)~\cite{Lewandowski2020}
\bea
\Pi(\bm{q},\omega)&=&g\sum_{\bm{k}}\sum_{s,s^{\prime}} f_{s,\bm{k}}\left[
\frac{ F^{ss^{\prime}}_{\bm{k},\bm{k}-\bm{q}} }{E_{s,
\bm{k}}- E_{s^{\prime},\bm{k}-\bm{q}} -\omega -i0 }
\right. \nn
\\
&&
\left.
-\frac{F^{s^{\prime}s}_{\bm{k}+\bm{q},\bm{k}}}{E_{s^{\prime},\bm{k}+\bm{q}}- E_{s,\bm{k}}-\omega -i0}
\right]~.
\label{Pi_Eq2}
\eea
{Here, the total polarization function has two  contributions, $\Pi_{\rm intra}(\bm{q},\omega)$ for $s=s^{\prime}$ summation, and $\Pi_{\rm inter}(\bm{q},\omega)$ for $s\ne s^{\prime}$ summation.  }
We define $\Delta E_{\bm{k},\bm{k}^{\prime}}\equiv (E_{s,\bm{k}}-E_{s,\bm{k}^{\prime}}$).
We can expand eigenvalues  up to ${\mathcal O}(q^3)$ as,
\bea
E_{s,\bm{k}\pm{\bm{q}}}&\approx& E_{s,\bm{k}} \pm {q}_a{v_{s,\bm{k}}^a} + \frac{q_aq_b}{2}\frac{\partial^2 E_{s,\bm{k}}}{\partial k_a \partial k_b}\nn
\\
&\pm& \frac{q_aq_bq_c}{6}\frac{\partial^3 E_{s,\bm{k}}}{\partial k_a \partial k_b \partial k_c}~,
\eea
where $v_{s,\bm{k}}=\nabla_{\bm{k}}E_{s,\bm{k}}$.
For substantially small $q$, the energy associated with the intraband transitions $\Delta E_{\bm{k},\bm{k}^{\prime}}$ 
are always smaller than $\omega$. Besides, the interband transition energy will be larger than $\omega$. So, in small ${\bm q}$ limit, we can expand the intraband polarization function in powers of $1/\omega$~\cite{giuliani2005quantum} 
\bea
\Pi_{\rm intra}&=&g\sum_{\bm{k}}{f_{s,\bm{k}}}\left[ 
\frac{ F^{ss}_{\bm{k},\bm{k}-\bm{q}} }{\Delta E_{\bm{k},\bm{k}-\bm{q}} -\omega}
-\frac{F^{ss}_{\bm{k}+\bm{q},\bm{k}}}{\Delta E_{\bm{k}+\bm{q},\bm{k}}-\omega} \right], \nn
\\
&=& g\sum_{\bm{k}}\frac{f_{s,\bm{k}}}{\omega}\left[ 
\frac{F^{ss}_{\bm{k}+\bm{q},\bm{k}}}{\left(1- \frac{ \Delta E_{\bm{k}+\bm{q},\bm{k}} }{\omega} \right)}
-\frac{ F^{ss}_{\bm{k},\bm{k}-\bm{q}} }{ \left(1- \frac{ \Delta E_{\bm{k},\bm{k}-\bm{q}} }{\omega} \right) }\right], \nn
\\
&\approx& \frac{A_1(\bm{q})}{\omega} + \frac{A_2(\bm{q})}{\omega^2}+ \frac{A_3(\bm{q})}{\omega^3}+\dots~.
\label{Pi_Eq4}
\eea

\section{Derivation of Eq.~\eqref{Plasmon_disp}}\label{Eq.12_derivation}
We perturbatively solve Eq.~\eqref{Eq_roots} as
\bea
\label{perturvative solve}
\omega_p^2&=& V_{|\bm{q}|}^{(d)}A_2 + V_{|\bm{q}|}^{(d)}\omega_pA_2 + \frac{V_{|\bm{q}|}^{(d)}A_3}{\omega_p} \nn~,
\\
\omega_p^2&\approx& (\omega_p^0)^2 + V_{|\bm{q}|}^{(d)}A_1\omega_p^{0}
+\frac{V_{|\bm{q}|}^{(d)}A_3}{\omega_p^0} \nn~,
\\
\omega_p &\approx& \omega_p^{0}\left[ 1+ \frac{A_1}{\omega_p^0}+ \frac{V_{|\bm{q}|}^{(d)}A_3}{q(\omega_p^0)^3}  \right]^{1/2} \nn~,
\\
\omega_p(\bm{q})&\approx& \omega_p^{0} +\frac{1}{2}V_{|\bm{q}|}^{(d)}A_1 +\frac{A_3}{2A_2} +{\cal O}\left(A_3,A_1\right)^2~.
\eea

\section{Reciprocal interband plasmon in QWZ model in the presence of inversion symmetry}\label{Appendix_reciprocal_interband}
\begin{figure}[h!]
    \includegraphics[width=\linewidth]{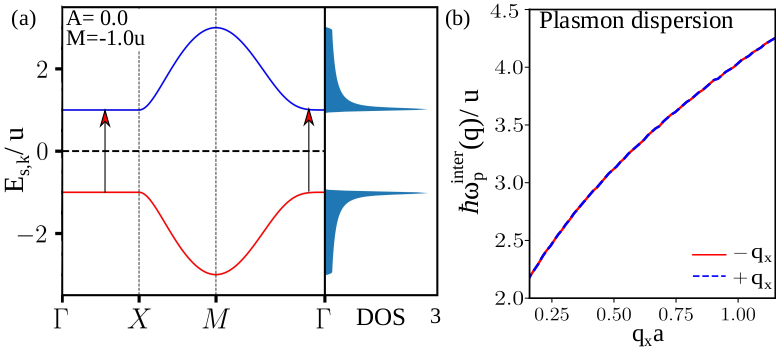}
    \caption{(a) Electronic band dispersion and DOS of the QWZ model with parameters $A=0$ and $M=-1$u. This Hamiltonian preserves ${\mathcal P}$ but breaks ${\mathcal T}$. (b) The interband plasmon dispersion for this system is obtained by solving for the roots of the dielectric function. This shows symmetric (or reciprocal) plasmon dispersion for $+\bm{q}$ and $-\bm{q}$ propagation in the presence of inversion symmetry.     
    }
    \label{fig_reciprocal_inter}
\end{figure}

This appendix conveys that interband plasmon will be reciprocal when the system has inversion symmetry but broken time-reversal symmetry. So, we set the parameters, $A=0$ and $M=-1$u in Eq.~\eqref{QWZ_model_Eq}. We calculate the band dispersion in fig.~\ref{fig_reciprocal_inter}(a). These bands are nearly flat over BZ, exhibiting VHS in the JDOS spectrum.  
However, due to the presence of inversion symmetry, ${\mathcal B}$ turns out as zero and leads to reciprocal interband mode as shown in Fig.~\ref{fig_reciprocal_inter}(b). 

\section{Symmetry properties of different geometric quantities}
In this section, we present the symmetry properties of various geometric quantities, like ${\mathcal Q}_{abc}$, ${\mathcal C}_{abc}$ and ${\mathcal B}_{abc}$ under ${\mathcal P}$ and ${\mathcal T}$ operation.
\subsection{Symmetry of ${\mathcal Q}_{abc}$}\label{symmetry_G_abc}
In the presence of ${\mathcal T}$ or ${\mathcal P}$ symmetry, the quantum metric obeys the relation~\cite{Hughes2019} $g^{ab}_{s}(-\bm{k})=g^{ab}_{s}(\bm{k})$. In presence of ${\cal P}$ or ${\cal T}$, $E_{s,\bm{k}}=E_{s,-\bm{k}}$. 
The Fermi function ($f_{s,\bm{k}}$) is also an even function of $\bm{k}$, i.e. $f_{s,\bm{k}}= f_{s,-\bm{k}}$ in presence of ${\mathcal P}$ or ${\mathcal T}$ because it is the only function of $E_{s,\bm{k}}$. As a result, change of the integration variable from $\bm{k}$ to $-\bm{k}$ in Eq.~\eqref{Eq_A1}, owing to the ${\mathcal T}$ or ${\mathcal P}$  symmetry, the ${\bm k}$- integrant of ${\mathcal Q}_{abc}$ becomes an odd function of $\bm{k}$ and vanishes as we sum over Brillouin zone. Thus, one need to break both ${\mathcal P}$ and ${\mathcal T}$ symmetries to get ${\mathcal Q}_{abc}$ non-zero.

\subsection{Symmetry of ${\mathcal C}_{abc}$}\label{symmetry_C_abc}
In the presence of ${\mathcal T}$ or ${\mathcal P}$, 
the Fermi function ($f_{s,\bm{k}}$) is also an even function of $\bm{k}$ i.e. $f_{s,\bm{k}}= f_{s,-\bm{k}}$. Now, either in presence of ${\cal P}$ or ${\cal T}$, the velocity operator, $v^{a}_{s,\bm{k}}=-v^{a}_{s,-\bm{k}}$. 
So, as we change the integration variable from $\bm{k}$ to $-\bm{k}$ in Eq.~\eqref{A3}, owing to the ${\mathcal T}$ or ${\mathcal P}$ symmetry, the overall $\bm{k}$- integrant of $C_{abc}$ becomes an odd function of $\bm{k}$, and vanishes as we sum over the whole Brillouin zone. Thus, one need to break both ${\mathcal P}$ and ${\mathcal T}$ symmetry to get ${\mathcal C}_{abc}$ non-zero.

\subsection{Symmetry of ${\mathcal B}_{abc}$}\label{symmetry_B_abc}
In main text, We have defined ${B}_{abc}=\sum_{\bm{k}}\sum_{s,s^{\prime}}f_{s,\bm{k}}F^{(3)}_{abc,ss^{\prime}}$, where $F^{(3)}_{abc,ss^{\prime}}={\rm Re}\left[ \langle u_{s,\bm{k}} | \partial_{{k}_a} u_{s^{\prime},\bm{k}}\rangle \langle \partial_{{k}_b}\partial_{{k}_c}u_{s^{\prime},\bm{k}}| u_{s,\bm{k}}\rangle \right]$. Now, let us investigate the inversion and time-reversal operations one by one.\\
\begin{figure*}[!t]
    \includegraphics[width=\linewidth]{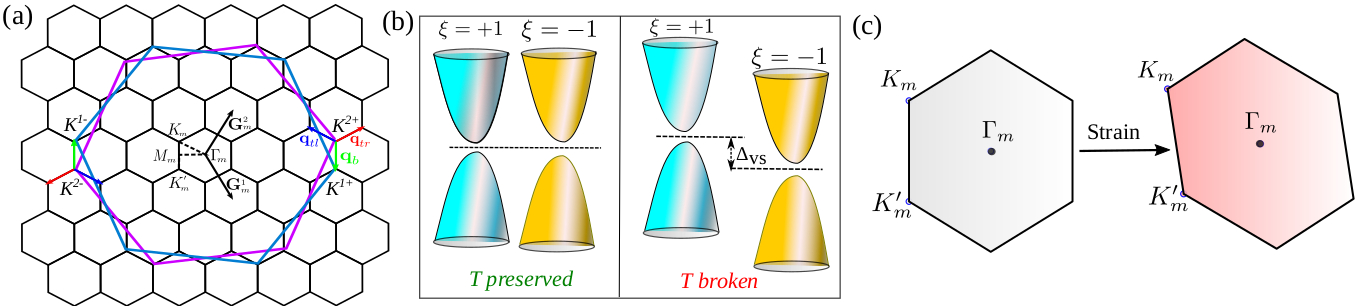}
    \caption{ (a) Twisted hexagonal graphene Brillouin zones (cyan and purple lines) and folded moir\'{e} mini Brillouin zones (black lines) of moir\'{e} superlattice. The red, green, and blue arrows represent the nearest neighbors between the two layers. The orientation of the arrows on the right-hand side represents $K$-valley ($\xi=+1$) and the left-hand side represents $K^{\prime}$ valley ($\xi=-1$). (b) Schematic illustration of time-reversal symmetry breaking through finite valley splitting between the $K$ and the $K^{\prime}$ valley. Panel (c) represents the effect of uniaxial strain on the moir\'{e} Brillouin zone. { Strain breaks all spatial symmetries of twisted bilayer graphene except C$_{2z}$.}     
    }
    \label{moire_BZ}
\end{figure*}
\noindent\textbf{Parity (${\mathcal P}$):}
In presence of parity symmetry, the Bloch Hamiltonian ${\mathcal H}_{\bm{k}}$ satisfy; ${\mathcal P}{\mathcal H}_{\bm{k}}{\cal P}^{-1}={\cal H}_{-\bm{k}}$. The eigenfunction changes as ${\cal P}u_{s,\bm{k}}(\bm{r})=u_{s,-\bm{k}}(-\bm{r})$. So, under parity operations, we get
\bea
F^{(3)}_{abc,ss^{\prime}}(\bm{k})
&=&{\rm Re}\left[ \langle u_{s,-\bm{k}} | \partial_{{k}_a} u_{s^{\prime},-\bm{k}}\rangle \langle \partial_{{k}_b}\partial_{{k}_c}u_{s^{\prime},-\bm{k}}| u_{s,-\bm{k}}\rangle \right]~,\nn
\\
&=& -F^{(3)}_{abc,ss^{\prime}}(-\bm{k}).
\eea
The Fermi function ($f_{s,\bm{k}}$) is also an even function of $\bm{k}$ i.e. $f_{s,\bm{k}}= f_{s,-\bm{k}}$ under parity operation. So, owing to ${\cal P}$ symmetry, the overall $\bm{k}$- integrant of ${\cal B}_{abc}$ is an odd function of $\bm{k}$. Therefore, in presence of ${\cal P}$, ${\cal B}_{abc}$ turns out as zero.\\

\noindent\textbf{Time-reversal (${\mathcal T}$):}
For a spinless system, the time-reversal symmetry operator ($\mathcal T$) can be expressed by complex conjugation operation $\mathcal T={\mathcal K}$. In presence of time-reversal symmetry, the Bloch  Hamiltonian; ${\cal H}_{\bm{k}}$ satisfy; ${\mathcal T}{\mathcal H}_{\bm{k}}{\cal T}^{-1}={\cal H}_{-\bm{k}}$. The eigenfunction changes as ${\cal T}u_{s,\bm{k}}(\bm{r})=u_{s,-\bm{k}}^{*}(\bm{r})$. So, under time-reversal transformation, the interband Berry connection ${\cal R}^{a}_{ss^{\prime}}(\bm{k})$ ($=\langle u_{s, \bm{k}}|\partial_{k_a}u_{s^{\prime},\bm{k}}\rangle$) modifies as,
\bea
{\cal R}_{ss^{\prime}}^{a}(\bm{k})
&=& \langle u_{s,-\bm{k}}^{*}|\partial_{k_a} u_{s^{\prime},-\bm{k}}^{*}\rangle~,\nn
\\
&=& -\langle u_{s^{\prime},-\bm{k}}|\partial_{k_a} u_{s,-\bm{k}}\rangle~,\nn
\\
&=& {\cal R}^{a}_{s^{\prime}s}(-\bm{k})
\eea
Therefore, under time-reversal operations, we have
\small{
\bea
F^{(3)}_{abc,ss^{\prime}}(\bm{k})
&=& {\rm Re}\left[ {\cal R}_{s^{\prime}s}(-\bm{k}) \langle \partial_{{k}_b}\partial_{{k}_c}u_{s^{\prime},-\bm{k}}^{*}| u_{s,-\bm{k}}^{*}\rangle \right]~,\nn
\\
&=& {\rm Re}\left[\langle \partial_{{k}_a} u_{s^{\prime},-\bm{k}}|u_{s,-\bm{k}}\rangle \langle u_{s,-\bm{k}}|\partial_{{k}_b}\partial_{{k}_c}u_{s^{\prime},-\bm{k}}\rangle \right]~,\nn
\\
&=& -{\rm Re}\left[\langle u_{s,-\bm{k}} |\partial_{-{k}_a} u_{s^{\prime},-\bm{k}}\rangle^{*} \langle \partial_{-{k}_b}\partial_{-{k}_c}u_{s^{\prime},-\bm{k}}|u_{s,-\bm{k}}\rangle^{*} \right]~,\nn
\\
&=& -F^{(3)}_{abc,ss^{\prime}}(-\bm{k})~.\nn
\eea
}
\normalsize
So, owing to ${\cal T}$ symmetry, the overall $\bm{k}$- integrant of ${\cal B}_{abc}$ is an odd function of $\bm{k}$. Therefore, in presence of ${\cal P}$, ${\cal B}_{abc}$ turns out as zero. Therefore, to get the non-zero value of ${\mathcal B}_{abc}$, we need to break both ${\cal P}$ and ${\cal T}$ simultaneously.  

\subsection{Symmetry of ${\cal A}_{ab}$}\label{symmetry_A}
In Eq.~\eqref{def_A}, We have defined ${\cal A}_{ab}=g\sum_{\bm{k}}\sum_{s, s^{\prime}}^{s^{\prime}\ne s} f_{s,\bm{k}}{\cal R}^{a}_{ss^{\prime}}{\cal R}^{b}_{s^{\prime}s}$. In the presence of inversion symmetry, we have ${\cal R}^{a}_{ss^{\prime}}(\bm{k})=-{\cal R}^{a}_{ss^{\prime}}(-\bm{k})$, and in presence of TRS we have, ${\cal R}^{a}_{ss^{\prime}}(\bm{k})={\cal R}^{a}_{s^{\prime}s}(-\bm{k})$. The Fermi-distribution function also follows, $f_{s,\bm{k}}=f_{s,-\bm{k}}$ in presece of ${\cal P}$ or ${\cal T}$. So, owing to ${\cal P}$ or ${\cal T}$ symmetry, the overall $\bm{k}$- integrant of ${\cal A}_{ab}$ is an even function of $\bm{k}$ and turns out as non-zero under $\bm{k}$ integrations.

\section{Continuum model Hamiltonian for Twisted Bilayer Graphene}\label{Appendix_continuum_model}
The successful formulation of a rigorous TBG band theory for small twist angle $\theta$ can be obtained from the continuum model originally proposed by {\textit{Santos-Peres-Neto}~\cite{Santos2007G} and later also reported by} \textit{Bistritzer-MacDonald}~\cite{Bistritzer2011, Andrei2020}. In this section, we discuss the construction of the TBG model Hamiltonian. For the band structure calculation, we assume that the top bilayer ($l= 1$) is rotated by an angle $-\theta/2$ and the bottom bilayer ($l= 2$) is rotated by an angle $\theta/2$. The modified reciprocal lattice vectors are ${\bm b}_i^{(l)} = R(\mp \theta/2) {\bm a}_i^*$, where $R$ is the two-dimensional rotation matrix. The reciprocal primitive vectors , follow the relation ${\bm a}^*_i\cdot {\bm a}_j=2\pi\delta_{i,j}$. Here the primitive lattice real space vectors are $\bm{a}_1=a(1,0)$, and $\bm{a}_2=a(1/2,\sqrt{3}/2)$ with $a=2.46~\AA$ being the lattice constant.  The reciprocal lattice vectors for the moir\'{e} superlattices are given as ${\bm g}_i=b^{(1)}_{i}-b^{(2)}_{i}$.
The effective low-energy Hamiltonian for valley $\xi=\pm$ can be expressed as
\be \label{BM_s}
H_{\xi}({\bf k})=
\begin{pmatrix}
h_{-\theta/2,\xi}+\Delta_I \sigma_z & T_\xi({\bf r}) \\
T^\dagger_\xi ({\bf r})& h_{\theta/2,\xi} +\Delta_I\sigma_z
\end{pmatrix}
\ee
$h_{\theta.\xi}= -\hbar v_F R(\theta) (k-D_{l,\xi}).[\xi\sigma_x,\sigma_y]$ where $k_l$ are the momentum space location of the Dirac points of $l^{th}$ layer. $\xi=\pm 1$ represents the $K$ and $K^\prime$ valley respectively. $v_F$ is the Fermi velocity of our system. We choose $\hbar v_F=$ 5.96 eV.\AA~\cite{KTLaw2022} for our continuum model calculations. $\sigma_i$ with ($i=1-3$) is the sublattice Pauli
matrices of the single-layer graphene Hamiltonian. $\Delta_I$  dictates the {inversion breaking sublattice potential difference that promotes a gap at $K$ and $K^{\prime}$,  similar to that in graphene-hexagonal Boron Nitride system}~\cite{Wang2016G}. This gap can be introduced in the system in the presence of the finite external electric field, substrate effect, etc. The inter-layer coupling Hamiltonian in the $BM$ model is given by
\small 
\be \label{T_j}
T_\xi(\bm{r}) = 
\begin{pmatrix}
u_0 & u_1 \\
u_1& u_0
\end{pmatrix}
+
\begin{pmatrix}
u_0 & u_1 e^{-i \xi \frac{2\pi}{3}} \\
u_1e^{i \xi\frac{2\pi}{3}} &  u_0
\end{pmatrix}e^{i\xi \bm{g}_1\cdot\bm{r}}
+ 
\ee
\be
\begin{pmatrix}
u_0 & u_1 e^{i \xi\frac{2\pi}{3}} \\
u_1e^{i \xi\frac{2\pi}{3}}  &  u_0 
\end{pmatrix}e^{i\xi(\bm{g}_1+\bm{g}_2)\cdot\bm{r}}. \nn
\ee
\normalsize

The diagonal and off-diagonal hoppings are considered $u_0=$79.7 meV and $u_1=$ 97.5 meV~\cite{Koshino2018M}. The matrix dimension used in all of our calculations is equivalent to moir\'{e} periodicity 3G where G is the reciprocal lattice vector. The moir\'{e} BZ, together with the high symmetry points, are shown in Fig.~\ref{moire_BZ}(a). To break the ${\cal T}$ symmetry of the system, a finite valley splitting ($\Delta_{vs}$) is included between the two valleys as schematically shown in Fig.~\ref{moire_BZ}(b). 

\begin{figure*}[t!]
    \includegraphics[width=0.8\linewidth]{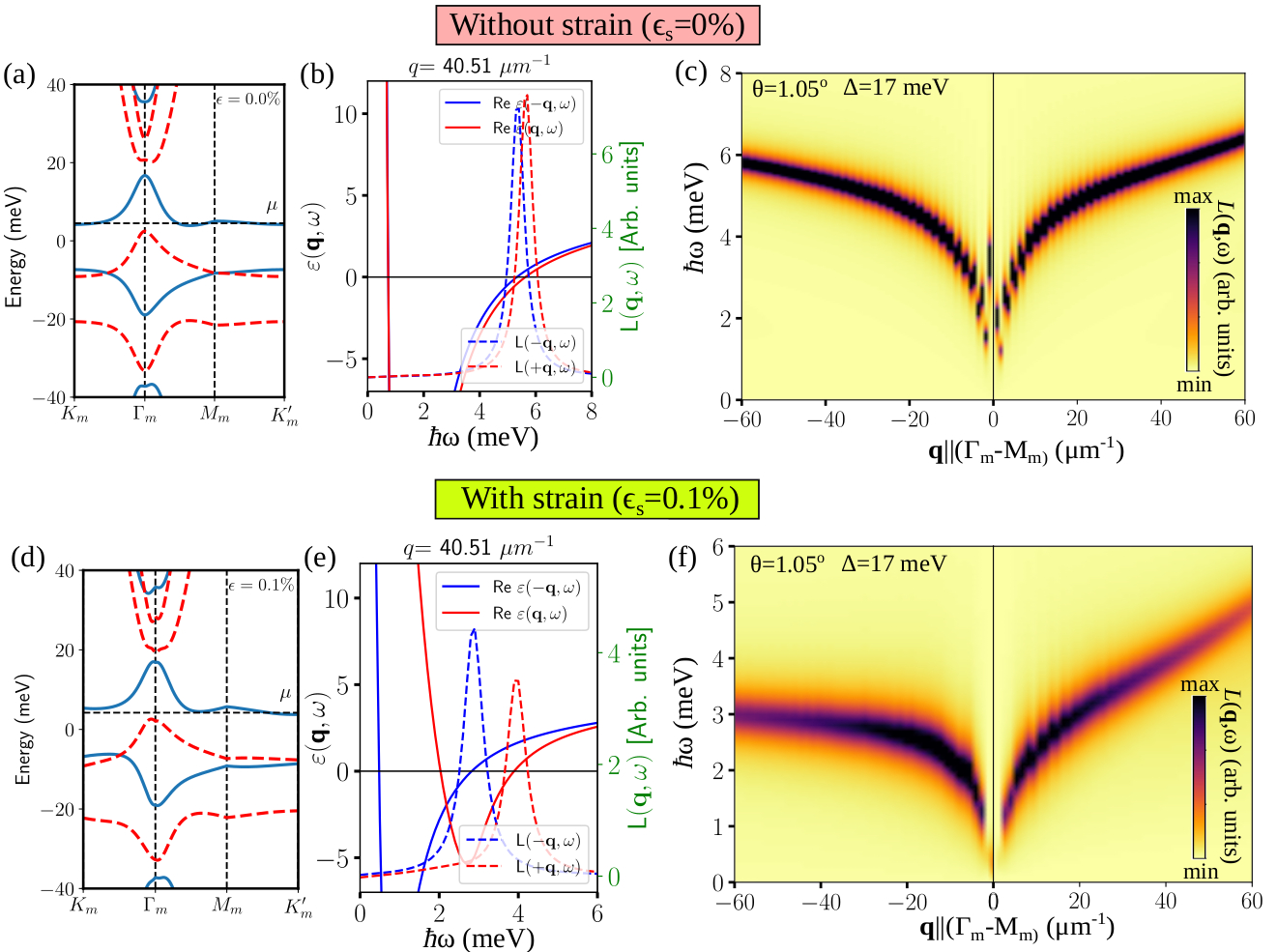}
 \caption{{Band dispersion of magic angle TBG (a) without strain ($\epsilon_s$) and (d) with $0.1$\% uniaxial strain. We have introduced a staggered potential $\Delta$=17 meV to mimic hBN-aligned TBG. The black dashed horizontal line represents the position of the chemical potential.  To set chemical potential ($\mu$) at flat band region, we choose $\mu=$3.5 meV and 4.5 meV for $\epsilon_s=0\%$ and without strain cases, respectively. RPA dielectric function (for +\textbf{$q$} and -\textbf{$q$} wavevector) with finite valley splitting ($\sim$16 meV) $\epsilon_s$=0 and $\epsilon_s$=0.1\% are shown in (b) and (e) respectively. We observe finite nonreciprocity due to the absence of parity and time-reversal symmetry, even in the absence of strain in the system. (c) and (f) show the colormap of the loss function in $\textbf{q}-\omega$ space. The inclusion of strain enhances the magnitude of nonreciprocity of the plasmon dispersion.}}
    \label{strain_intra}
\end{figure*}
\section{Implementation of strain in moir\'{e} Hamiltonian}
\label{Matrix_strain}

In this section, we describe the details of the strain implementation in the TBG continuum model Hamiltonian. Strain generally appears in 2D moir\'e systems while fabricating on the h-BN substrate. In the case of TBG, a strain may appear on both of the layers when sandwiched between the substrates ~\cite{bi_PRB2019_designing,pantaleon2021T}. However, for simplicity, in our calculation, we assume uniaxial strain ($\mathcal E$) acts only on a single (bottom) layer~\cite{he_NC2020} [see Fig.~\ref{moire_BZ}(c)]. 
In the presence of strain, the primitive lattice vectors and hence the reciprocal lattice vectors get distorted. For a given strain matrix ${\mathcal E}$, (which satisfies ${\mathcal E}^T={\mathcal E}$ with $T$ denoting the transpose), the real space vectors distort as ${\bm r} \to (1 + {\mathcal E}) {\bm r}$ and the reciprocal vectors as ${\bm k} \to (1 - {\mathcal E}^T) {\bm k}$. We obtain the strained moir\'{e} lattice vectors as ${\bm G}_{m}^i = {\bm b}_i^{(1)} - {\bm b}_i^{(2)}$ where ${\bm b}_{i}^{(l)}$ {represents the modified reciprocal lattice vectors due to combined effect of rotation and strain.} 
The impact of strain on the nonreciprocity of the TBG plasmonic modes is summarized in section~\ref{Sec-V}.

To calculate the strained band structure, we consider the uni-axial strain of strength ${\mathcal E}$ at an angle $\phi$ relative to the zigzag direction as~\cite{bi_PRB2019_designing,pantaleon2021T, sonia_PRBL2021_twist}
\be \label{strain}
{\mathcal E} = \epsilon_s 
\begin{pmatrix}
-\cos^2 \phi + \nu \sin^2 \phi & (1+\nu ) \sin \phi \cos \phi \\
(1+\nu) \sin \phi \cos \phi & -\sin^2 \phi + \nu \cos^2 \phi
\end{pmatrix}~.
\ee
Equation~\ref {strain} represents the system when it is more stretched in one direction, and less stretched in the perpendicular direction.
With this strain matrix, $\epsilon_s=0.1\%$ strain (with $\phi=0$) is used to calculate the electronic structure and plasmon dispersion of TBG. 
In the presence of strain, the Dirac Hamiltonian of Eq.~\eqref{BM_s} modifies to
\be 
h_{\mp \theta/2,\xi } =  \hbar v_F {\mathcal R}(\mp \theta/2)~[(\mathbb{I} + {\mathcal E}^T)  ] ({\bm k} - {\bm D}_{\xi}) \cdot (\xi \sigma_x, \sigma_y) 
\ee
Here, the strain matrix operates over the position of the twisted Dirac points given by
\be
{\bm D}_\xi = (\mathbb{I} - {\mathcal E}^{T}) {\bm K}^i_\xi - \xi {\bm A}~,
\ee
with ${\bm A}$ representing the gauge field that has the dimension of reciprocal lattice vector. The sublattice potential difference is included in the Hamiltonian through a gap of $\sim$17 meV. The appearance of the gauge field can be attributed to the fact that the strain causes the inter-atomic distance in each layer to become different in different directions. This results in the difference of hopping parameters which displaces the Dirac point from its original position. The gauge potential ${\bm A}$ in terms of the elements of the strain matrix is given by
\be 
{\bm A} = \dfrac{\sqrt{3}}{{2a}} \beta ({\mathcal E}_{xx} - {\mathcal E}_{yy} , -2 {\mathcal E}_{xy})~.
\ee
Here, $\beta=1.57$ and ${\mathcal E}_{ij}$ are the elements of the strain matrix [see \eqref{strain}].

Strain also modifies the lattice vectors and, consequently, the hopping matrices and the hopping vectors. We calculate the strained moir\'e vectors starting from un-rotated and un-strained lattice vectors. Following Refs.~\cite{he_NC2020}, we obtain the lattice vectors using ${\bm G}_{m}^{1, {\rm st}}=R_{-\frac{\theta}{2}} (1-{\mathcal E}^T){\bm b}_1 - R_{\frac{\theta}{2}} {\bm b}_1$ and ${\bm G}_{m}^{2, {\rm st}}=R_{-\frac{\theta}{2}} (1-{\mathcal E}^T){\bm b}_2 - R_{\frac{\theta}{2}} {\bm b}_2$ which yields
\begin{widetext}
\begin{subequations}
\bea
{\bm G}_{m}^{1, \rm st} &=&\dfrac{k_\theta}{4} \Big( 2\sqrt{3} - 3{\mathcal E}_{xy} - \sqrt{3} {\mathcal E}_{yy} - (3{\mathcal E}_{xx} +\sqrt{3} {\mathcal E}_{xy})\cot \frac{\theta}{2},~ -6 + 3{\mathcal E}_{xx} + \sqrt{3} {\mathcal E}_{xy} -(3 {\mathcal E}_{xy} + \sqrt{3} {\mathcal E}_{yy})\cot \frac{\theta}{2} \Big),~~~
\\
{\bm G}_{m}^{2, \rm st}
&=&\dfrac{k_\theta}{4} \Big( 2\sqrt{3} + 3{\mathcal E}_{xy} - \sqrt{3} {\mathcal E}_{yy} + (3{\mathcal E}_{xx} -\sqrt{3} {\mathcal E}_{xy})\cot \frac{\theta}{2},~
 6 - 3{\mathcal E}_{xx} + \sqrt{3} {\mathcal E}_{xy} +(3 {\mathcal E}_{xy} - \sqrt{3} {\mathcal E}_{yy})\cot \frac{\theta}{2} \Big).~~~
\eea 
\end{subequations}
\end{widetext}

{Implementing the strained Hamiltonian, we have calculated dielectric function and loss function spectra as shown in Fig.~\ref{strain_intra}. The band dispersions without and with  strain ($\epsilon_s=0.1\%$) are shown in Fig.~\ref{strain_intra} (a) and (d), respectively. Effect of strain makes the band-geometric effect more pronunced~\cite{pantaleon2021T,Arpit2021S}.
Even, without strain, we also observed nonreciprocal intraband plasmon in TBG as shown in Fig.~\ref{strain_intra} (b) and (c). 
We have found that incorporation of strain, enhanced the nonreciprocity in TBG for long-wavelength limit. This is shown in Fig.~\ref{strain_intra}. } 
\normalsize
\section{Derivation of Eq.~\eqref{Interband_overlap_Eq} }\label{appendix_H}
We expand the interband coherence term $|\langle u_{s,\bm{k}}| u_{s^{\prime},\bm{k}+\bm{q}}\rangle|^2$ up to third order of $q$. First, we Taylor expand for small $q$ as
\bea
|u_{s^{\prime},\bm{k}+\bm{q}}\rangle= |u_{s^{\prime},\bm{k}}\rangle+q_a|\partial_{k_a}u_{s^{\prime},\bm{k}}\rangle + \frac{q_aq_b}{2}|\partial_{k_a}\partial_{k_b}u_{s^{\prime},\bm{k}}\rangle \nonumber \\ 
+\frac{q_aq_bq_c}{6}|\partial_{k_a}\partial_{k_b}\partial_{k_c}u_{s^{\prime},\bm{k}}\rangle +{\cal O}(q^4)~~~.
\eea
The overlap between two states is calculated as,
 \bea
\langle u_{s,\bm{k}}|u_{s^{\prime},\bm{k}+\bm{q}}\rangle=q_a\langle u_{s,\bm{k}}|\partial_{k_a}u_{s^{\prime},\bm{k}}\rangle + \frac{q_aq_b}{2}\langle u_{s,\bm{k}}|\partial_{k_a}\partial_{k_b}u_{s^{\prime},\bm{k}}\rangle \nonumber \\ +\frac{q_aq_bq_c}{6}\langle u_{s,\bm{k}}|\partial_{k_a}\partial_{k_b}\partial_{k_c}u_{s^{\prime},\bm{k}}\rangle+{\cal O}(q^4)~~~~~~.
\eea   

The overall interband coherence term is computed as,
\bea
F^{ss^{\prime}}_{\bm{k},\bm{k}+\bm{q}}&=&\langle u_{s,\bm{k}}|u_{s^{\prime},\bm{k}+\bm{q}}\rangle\langle u_{s,\bm{k}}|u_{s^{\prime},\bm{k}+\bm{q}}\rangle^{*}\nn
\\
&=&q_aq_bF^{(2)}_{ab, ss^{\prime}}+q_aq_bq_cF^{(3)}_{abc, ss^{\prime}} +{\mathcal O}(q^4)~.
\eea
Here the expansion coefficients are given by
\bea
F^{(2)}_{ab,ss^{\prime}}&=&\langle u_{s,\bm{k}}|\partial_{k_a}u_{s^{\prime},\bm{k}}\rangle\langle\partial_{k_b}u_{s^{\prime},\bm{k}}|u_{s,\bm{k}}\rangle~,
\\
F^{(3)}_{abc,ss^{\prime}}&=&{\rm Re}\left[ \langle u_{s,\bm{k}} | \partial_{{k}_a} u_{s^{\prime},\bm{k}}\rangle \langle \partial_{{k}_b}\partial_{{k}_c}u_{s^{\prime},\bm{k}}| u_{s,\bm{k}}\rangle \right]~.
\eea
\twocolumngrid

\section{Nonreciprocal plasmon in 1D bipartiate lattice model}\label{Appendix-K}
\begin{figure}[t!]
    \includegraphics[width=\linewidth]{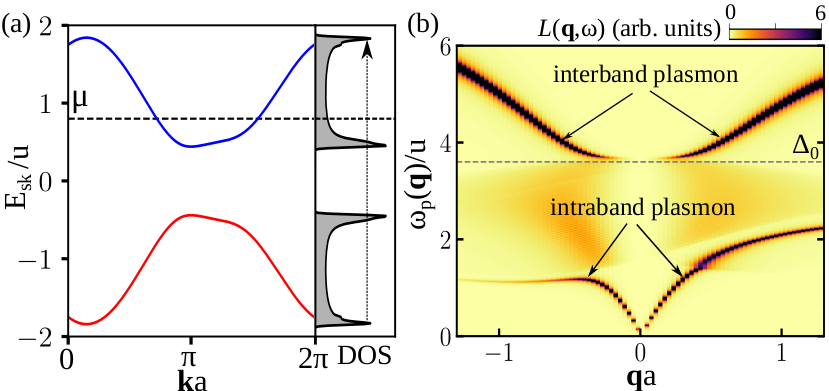}
    \caption{ 
	(a) The electronic band dispersion and density of states of the 1D magnetic bipartite lattice model with broken parity and time-reversal symmetry. We have used the parameters $v=0.4u$, $w=0.5u$, and $\phi=1$. (b) Colormap of the RPA loss function spectrum 
    showing the nonreciprocal intraband and interband plasmon. The interband plasmon starts from $\Delta_0$, which is the energy difference between two VHS as marked in panel (a).
    We also used the lateral confinement $R=0.2a$ and dimensionless constant $e^2/(4\pi\kappa\epsilon_0ua)=18$ in Eq.~\eqref{Eq_Vq}. 
    }
    \label{1D_model_fig}
\end{figure}
To illustrate the possibility of intrinsic nonreciprocal bulk plasmons, we consider another 1D model system of a magnetic bipartite lattice described in Ref.~\cite{Justin_song2022}. The intra- and inter-cell nearest-neighbor hopping amplitudes are $u$ and $v$, respectively. In addition, a complex third-neighbor hopping amplitude $w=|w|e^{i\phi}$, with phase $\phi$ has been considered here. The on-site energy is set to zero. The $k-$dependent tight-binding Hamiltonian, in basis $\{|A\rangle,~|B\rangle\}$, is given as
\bea
H_{\bm k}=\begin{pmatrix} 0& J^{*}({k}) \\
J({k})& 0\\
\end{pmatrix}~.
\eea
with $J({k})=ue^{-ika/2}+ve^{ika/2}+|w|e^{-i3ka/2}e^{i\phi}$, where $a$ is the lattice constant. This model described two bands with eigenvalues $E_{s\bm{k}}=s|J(k)|$, with eigenvector
\bea
|\bm{k},s\rangle =\frac{1}{\sqrt{2}}\begin{pmatrix}1 \\ se^{i\theta_{k}}
\end{pmatrix}
\eea
where $s=-1~(+1)$ denotes valence (conduction) bands and $\theta_k=arg[J(k)]$. The band dispersion and DOS spectrum are shown in Fig.~\ref{1D_model_fig}.

Considering that the spin-degree of freedom is not involved, the inversion (${\mathcal P}$) and time-reversal (${\mathcal T}$) operator is defined in the basis of $H_k$ as ${\mathcal P}\to \sigma_x$ and ${\mathcal T}\to K$, where ${\mathcal K}$ is complex conjugation. This model breaks both ${\mathcal P}$ and ${\mathcal T}$ for non-zero value of phase $\phi$, because $\bm{A} H_{\bm k}{\bm A}^{-1}\ne H_{-\bm k}$, for ${\bm A}= {\cal P}, {\cal T}$.

Next, we numerically compute the total electron polarization function and RPA dynamical dielectric function by using Eq.~\eqref{Pi_Eq1} and Eq.~\eqref{Eq_epsilon}. From the color plot of the loss function spectrum in Fig.~\ref{1D_model_fig}, we have found gapless nonreciprocal intraband mode and as well as gapped interband mode starting from $\Delta_0$ [the energy difference between two VHSs in Fig.~\ref{1D_model_fig} (a)]. This model provides another example where both \textit{intrinsic} nonreciprocal intraband and interband plasmon can be observed. \\

{
\section{Derivation of Eq.~\eqref{Pi_inter_expansion} }\label{appendix_interband_Pi}
We calculate the interband polarization function starting from Eq.~\eqref{Pi_Eq2} as
\bea
\Pi_{\rm inter}&=&g\sum_{\bm{k}}\sum_{s, s^{\prime}}^{s^{\prime}\ne s} f_{s,\bm{k}}\left[ 
\frac{ F^{ss^{\prime}}_{\bm{k},\bm{k}-\bm{q}} }{\Delta E_{\bm{k},\bm{k}-\bm{q}}^{ss^{\prime}} -\omega}
-\frac{F^{s^{\prime}s}_{\bm{k}+\bm{q},\bm{k}}}{\Delta E_{\bm{k}+\bm{q},\bm{k}}^{s^{\prime}s}-\omega} \right]~, \nn
\\
\label{Pi_inter_expansion2}
\eea
where $\Delta E_{\bm{k},\bm{k}^{\prime}}^{ss^{\prime}}\equiv (E_{s,\bm{k}}-E_{s,\bm{k}^{\prime}}$). We consider that a pair of bands are nearly flat over BZ and $E_{s^{\prime},\bm{k}+\bm{q}}- E_{s,\bm{k}}\approx \Delta_0$, where $\Delta_0$ denotes peak in JDOS. So, we can approximately calculate the interband polarization function as
\bea
\Pi_{\rm inter} &\approx& \frac{2g}{(\omega^2-\Delta_0^2)}\sum_{\bm{k}}\sum_{s,s^{\prime}}^{s^{\prime}\ne s} f_{s,\bm{k}}\left[ q_aq_bF^{(2)}_{ab,ss^{\prime}}\Delta_0 
\right.\nn
\\
&&
\left.
+\omega q_aq_bq_c F^{(3)}_{abc,ss^{\prime}} \right]~,\nn
\\
&\approx& \frac{2}{(\omega^2-\Delta_0^2)}\left[q_aq_b{\cal A}_{ab}\Delta_0 + q_aq_bq_c{\cal B}_{abc}\omega\right]~.
\eea
Here, ${\cal B}_{abc}$ and ${\cal A}_{ab}$ are defined in main text.
}

\section{Connection of $F^{(3)}_{abc,ss^{\prime}}$ with metric connection}\label{Gamma_mp}
{
The metric connection ${\Gamma}^{abc}_{s^{\prime}s}$ is a third-rank tensor. It is defined as the real part of the quantum geometric connections, ${\cal 
 G}^{abc}_{s^{\prime}s}$ as,
 \be
{\Gamma}^{abc}_{s^{\prime}s}~=~ {\rm Re}\left[{\cal 
 G}^{abc}_{s^{\prime}s} \right]~=~ {\rm Re}\left[{\cal R}^{a}_{ss^{\prime}}{\cal D}^b_{s^{\prime}s}{\cal R}^c_{s^{\prime}s}\right]~.
\label{Eq_L1}
\ee
Here,  ${\cal D}^b_{s^{\prime}s}=\partial_{k_b} -i({\cal R}^{b}_{s^{\prime}s^{\prime}}-{\cal R}^{b}_{ss})$ is the covariant derivative, and ${\cal R}^a_{ss^{\prime}}=i\langle u_{s,\bm{k}}|\partial_{k_a}u_{s^{\prime},\bm{k}}\rangle$ denotes Berry connection~\cite{Agarwal2022A,AhnPRX2020}.
In this appendix, we will show that the third rank tensor $F^{(3)}_{abc,ss^{\prime}}={\rm Re}\left[ \langle u_{s,\bm{k}} | \partial_{{k}_a} u_{s^{\prime},\bm{k}}\rangle \langle \partial_{{k}_b}\partial_{{k}_c}u_{s^{\prime},\bm{k}}| u_{s,\bm{k}}\rangle \right]$ is identical to with the metric connection $\Gamma^{abc}_{s^{\prime }s}$, for $a=b=c$. }
We can simplify Eq.~\eqref{Eq_L1} as
\begin{widetext}
\bea
&&{\cal R}^{a}_{ss^{\prime}}{\cal D}^b_{s^{\prime}s}{\cal R}^c_{s^{\prime}s}= i\langle u_s|\partial_{a}u_{s^{\prime}}\rangle\left[  \partial_b -i\left(i\langle u_{s^{\prime}}|\partial_{b}u_{s^{\prime}}\rangle -i\langle u_s|\partial_{b}u_s\rangle \right) \right]i\langle u_{s^{\prime}}|\partial_{c}u_s\rangle~, \nn
\\
&& = -\langle u_s|\partial_{a}u_{s^{\prime}}\rangle\partial_b\left[\langle u_{s^{\prime}}|\partial_{c}u_s\rangle\right] -\langle u_{s}|\partial_{a}u_{s^{\prime}}\rangle\langle u_{s^{\prime}}|\partial_{b}u_{s^{\prime}}\rangle\langle u_{s^{\prime}}|\partial_{c}u_s\rangle
+\langle u_s|\partial_{a}u_{s^{\prime}}\rangle\langle u_s|\partial_{b}u_s\rangle\langle u_{s^{\prime}}|\partial_{c}u_s\rangle~,\nn
\\
&& = -\langle u_s|\partial_{a}u_{s^{\prime}}\rangle\left[ \langle\partial_bu_{s^{\prime}}|\partial_cu_s\rangle +  \langle u_{s^{\prime}}|\partial_b\partial_cu_s\rangle \right] -\langle u_s|\partial_{a}u_{s^{\prime}}\rangle\langle u_{s^{\prime}}|\partial_{b}u_{s^{\prime}}\rangle\langle u_{s^{\prime}}|\partial_{c}u_s\rangle +\langle u_s|\partial_{a}u_{s^{\prime}}\rangle\langle u_s|\partial_{b}u_s\rangle\langle u_{s^{\prime}}|\partial_{c}u_s\rangle~.\nn
\\
\label{Eq_L2}
\eea
\end{widetext}
\onecolumngrid%
Here, we removed the $\bm{k}$ index in the subscripts.
Now starting from the identity, $\partial_b\partial_c\langle u_{s^{\prime}}|u_{s}\rangle=0$, we have
\bea
\langle u_{s^{\prime}}|\partial_b\partial_c u_{s}\rangle= -\langle \partial_b\partial_c u_{s^{\prime}}|u_{s}\rangle -\langle \partial_c u_{s^{\prime}}|\partial_b u_{s}\rangle - \langle \partial_b u_{s^{\prime}}|\partial_c u_{s}\rangle~.
\label{Eq_L3}
\eea
Substituting Eq.~\eqref{Eq_L3} in Eq.~\eqref{Eq_L2}, we have
\bea
{\cal R}^{a}_{ss^{\prime}}{\cal D}^b_{s^{\prime}s}{\cal R}^c_{s^{\prime}s}= \langle u_{s}|\partial_{a}u_{s^{\prime}}\rangle\langle \partial_b\partial_c u_{s^{\prime}}|u_{s}\rangle+ {\langle u_{s}|\partial_{a}u_{s^{\prime}}\rangle\left[\langle\partial_c u_{s^{\prime}}|\partial_{b}u_{s}\rangle - \langle u_{s^{\prime}}|\partial_{b}u_{s^{\prime}}\rangle\langle u_{s^{\prime}}|\partial_{c}u_{s}\rangle + \langle u_{s}|\partial_{b}u_{s}\rangle\langle u_{s^{\prime}}|\partial_{c}u_{s}\rangle \right]}~.\nn 
\\
\eea
We know that $\Gamma_{s^{\prime}s}^{abc}={\rm Re}\left[ {\cal R}^{a}_{ss^{\prime}}{\cal D}^b_{s^{\prime}s}{\cal R}^c_{s^{\prime}s} \right]$. This leads to
\bea
\Gamma_{s^{\prime}s}^{abc}=F^{(3)}_{abc,ss^{\prime}} +{\rm Re}\left[\langle u_{s}|\partial_{a}u_{s^{\prime}}\rangle\left[\langle\partial_c u_{s^{\prime}}|\partial_{b}u_{s}\rangle - \langle u_{s^{\prime}}|\partial_{b}u_{s^{\prime}}\rangle\langle u_{s^{\prime}}|\partial_{c}u_{s}\rangle + \langle u_{s}|\partial_{b}u_{s}\rangle\langle u_{s^{\prime}}|\partial_{c}u_{s}\rangle \right] \right]~.
\eea
Now by inserting the complete basis set to the term $\langle\partial_c u_{s^{\prime}}|\partial_{b}u_{s}\rangle$, we can arrive at, 
\bea
\Gamma_{s^{\prime}s}^{abc}&&=F^{(3)}_{abc,ss^{\prime}}+  {\rm Re}\left[\langle u_{s}|\partial_{a}u_{s^{\prime}}\rangle \langle \partial_c u_{s^{\prime}}|u_{s}\rangle\langle u_{s^{\prime}}|\partial_{b}u_{s^{\prime}}\rangle +  \langle u_{s}|\partial_{a}u_{s^{\prime}}\rangle \langle \partial_b u_{s^{\prime}}|u_{s}\rangle\langle u_{s^{\prime}}|\partial_{c}u_{s^{\prime}}\rangle \right]~. \\
&&=F^{(3)}_{abc,ss^{\prime}}+ Re[i{\cal R}^a_{ss^{\prime}}{\cal R}^{c}_{s^{\prime}s}{\cal R}^b_{s^{\prime}s^{\prime}} +i{\cal R}^a_{ss^{\prime}}{\cal R}^{b}_{s^{\prime}s}{\cal R}^c_{s^{\prime}s^{\prime}}]~.
\label{Eq_J7}
\eea
Here, the intraband Berry connection (${\cal R}^{a}_{ss}$) is purely a real quantity. For the longitudinal response, $a=b=c$, the product of the interband Berry connection terms, ${\cal R}_{ss^{\prime}}{\cal R}_{s^{\prime}s}$ is also a purely real quantity. Thus,  the second term of $\Gamma_{s^{\prime}s}^{abc}$ in Eq.~\eqref{Eq_J7} vanishes to zero. Thus, we can show that, ${\Gamma}^{aaa}_{s^{\prime}s}=F^{(3)}_{aaa,s^{\prime}s}$.
\twocolumngrid

\bibliography{ref}
\end{document}